\documentclass[a4paper,fleqn]{cas-dc}
\usepackage[utf8]{inputenc}
\usepackage[sort&compress,numbers]{natbib}
\usepackage{enumitem}
\usepackage[colorinlistoftodos]{todonotes}
\usepackage{color, soul}
\usepackage{multirow}
\usepackage{threeparttable}
\usepackage{caption}
\usepackage{xtab}
\usepackage[toc,page]{appendix}

\usepackage{microtype}

\def\tsc#1{\csdef{#1}{\textsc{\lowercase{#1}}\xspace}}
\tsc{WGM}
\tsc{QE}
\tsc{EP}
\tsc{PMS}
\tsc{BEC}
\tsc{DE}
\begin{document}
\let\WriteBookmarks\relax
\def\floatpagepagefraction{1}
\def\textpagefraction{.001}
\shortauthors{Tran et~al.}
\shorttitle{Assessing test artifact quality – A tertiary study}

\title[mode=title]{Assessing test artifact quality – A tertiary study}      

\author[1]{Huynh Khanh Vi Tran}[type=editor]
\ead{huynh.khanh.vi.tran@bth.se}

\address[1]{Blekinge Institute of Technology, Department of Software Engineering, SE-37179, Karlskrona, Sweden}

\author[1]{Michael Unterkalmsteiner}
\ead{michael.unterkalmsteiner@bth.se}

\author[1]{J\"urgen B\"orstler}
\ead{jurgen.borstler@bth.se}

\author[1]{Nauman bin Ali}
\ead{nauman.ali@bth.se}


\begin{abstract}
\textbf{Context:} Modern software development increasingly relies on software testing for an ever more frequent delivery of high quality software. This puts high demands on the quality of the central artifacts in software testing, test suites and test cases. 
\textbf{Objective:} We aim to develop a comprehensive model for capturing the dimensions of test case/suite quality, which are relevant for a variety of perspectives.
\textbf{Method:} We have carried out a systematic literature review to identify and analyze existing secondary studies on quality aspects of software testing artifacts.
\textbf{Results:} We identified 49 relevant secondary studies. Of these 49 studies, less than half did some form of quality appraisal of the included primary studies and only 3 took into account the quality of the primary study when synthesizing the results. We present an aggregation of the context dimensions and factors that can be used to characterize the environment in which the test case/suite quality is investigated. We also provide a comprehensive model of test case/suite quality with definitions for the quality attributes and measurements based on findings in the literature and ISO/IEC 25010:2011.
\textbf{Conclusion:} The test artifact quality model presented in the paper can be used to support test artifact quality assessment and improvement initiatives in practice. Furthermore, the model can also be used as a framework for documenting context characteristics to make research results more accessible for research and practice.

\end{abstract}


\maketitle

\section{Introduction}
Software development continues to use testing to ensure that we deliver high-quality software. The role of testing has become even more critical in the context of continuous software engineering~\cite{fitzgerald2017continuous}, as we are increasingly relying on testing to release reliable software more frequently.
In continuous software engineering organizations shorten the lead time from ideation to delivery by making features available to customers as soon as development is done.
This ambition of continuous delivery/deployment entails that organizations might deliver and deploy new version of a software which makes manual testing challenging.
Continuous software engineering, therefore, requires automated testing for quality assurance~\cite{shahin_continuous_2017}.

Software developers spend around one quarter of their effort on engineering tests~\cite{beller2015and}. It is therefore beneficial to quality assure test code and retaining the invested effort, i.e. assure that tests are effective. Test effectiveness (a test case/suite's ability to identify defects/faults in a system under test) has received a lot of attention in research, but it is only one of the multiple dimensions of test quality~\cite{bowes_how_2017}. 

Other dimensions include reliability of tests, such that when a test case fails it reveals a fault in the production system. Test code should thus be fault free to provide reliable verdicts on the production code~\cite{eck_understanding_2019}. Unfortunately, it is hard to develop bug-free test code~\cite{vahabzadeh2015empirical}. A further complication here is the flakiness of test cases~\cite{pinto_what_2020} which have non-deterministic output. Such unreliable tests take considerably more effort to troubleshoot and resolve.

In order to remain valuable, test cases must also co-evolve with production code~\cite{zaidman2011studying} and the current expected behavior of a SUT. The need for co-evolution requires that test cases are not only effective and fault free, but are also maintainable in the future. 

Proposals for defining test case quality~\cite{grano2020pinsa, huynh_khanh_test-case_2019, bowes_how_2017}, adapting existing quality models for testing~\cite{neukirchen_approach_2008} and proposals to organize software testing research~\cite{vegas_maturing_2009, ali_search_2019, engstrom_serp-test_2017} exist. However, there are no commonly agreed upon quality models, frameworks or taxonomies for the quality of test cases or test suites.

In the scope of this study, we are only focusing on test cases and test suites as the confidence in the findings of testing mainly depends on the quality of the test cases and test suites~\cite{bowes_how_2017, grano2020pinsa}. Therefore, it is essential to have the means to assess, monitor and maintain their quality. In this study, by test artifact, we refer to test cases, test suites, test scripts, test code, test specifications and natural language tests.

The contribution of this paper is a test artifact quality model. To develop this model, we have systematically investigated existing secondary studies of literature on software testing. We identified the quality attributes of a test case or test suite, and the context in which these attributes have been used, and how they can be measured. These attributes and associated measures were defined and compiled in the proposed test artifact quality model. 

For both researchers and practitioners, this comprehensive review presents an overview of the current state-of-research on test artifact quality. Furthermore, the quality model can be used to design and execute improvement initiatives targeting the quality of the test artifacts. 

The remainder of the paper is structured as follows: Section~\ref{sec:relatedWork} presents the related work and positions our contribution. Section~\ref{sec:method} details the design and procedures of the systematic literature review. Section~\ref{sec:results-analysis} presents the results and analysis for each of the research questions posed in the study. In Section~\ref{sec:discussion}, we discuss the selected findings related to quality attributes, measurements and the contexts in which they have been studied. Lastly, Section~\ref{sec:conclusion} concludes the paper.

\section{Related Work}\label{sec:relatedWork}
We discuss related work from three angles. First, we look at past research on test artifact quality. Then, we provide an overview of research that aims at classifying concepts in software testing. Finally, we look at the contributions made by other tertiary literature studies on software testing. We conclude this section by identifying the knowledge gaps that this study aims to fill.

\subsection{Test case and test suite quality}\label{sec:relatedWork_testArtifactQuality}
The question of test adequacy has been studied since the mid 1970s when Goodenough and Gerhard~\cite{goodenough1975toward} steered the research community's attention towards identifying criteria for evaluating test cases. Zhu et al.~\cite{zhu1997software} reviewed the literature of two decades (late 70s-late 90s) on test adequacy criteria, categorizing them into structural criteria (coverage of the test suite on the SUT), fault-based criteria (defect detection ability of the test suite) and error-based criteria (to what extent error-prone parts of the SUT are tested). Structural adequacy criteria include control flow, data flow and dependence coverage criteria, and are mostly based on the flow-graph model of a program. Structural adequacy criteria inform to what extent a test suite covers the control and data flows in a program. The higher the coverage, the better the quality but also the higher the cost to develop and maintain the test suite. Fault-based adequacy criteria measure the ability of a test suite to detect faults. Techniques to generate faults that could possibly be contained in software and should be found by adequate tests include error seeding, mutation testing, perturbation testing and the RELAY model. Error-based adequacy criteria use domain analysis to create equivalence partitions of the input/output space for the program. The domain analysis focuses thereby on particularly error-prone points in the program logic. The quality of a test suite is determined by the coverage of a programs partitions. 

Zhu et al.'s excellent review inspired other researchers to investigate structural test quality for specific application domains and programming paradigms, exemplified by the following studies. Kapfhammer and Soffa~\cite{kapfhammer2003family} define test suite adequacy criteria for database-driven applications. The goal of these criteria is to ensure that all database interaction associations involving relations, attributes, values, and records are exercised by the test suite. Lemos et al.~\cite{lemos2007control} propose structural testing criteria for aspect-oriented programs. They derive a control and data-flow model and create an aspect-oriented variable definition-use graph that leads to particular testing criteria that would have been missed by an approach for traditional programming paradigms. Finally, Pei et al.~\cite{pei_deepxplore_2019} propose a set of new testing criteria for deep neural networks (DNNs). Traditional criteria, such as statement and data-flow coverage, are not effective when testing DNNs since the system's logic is not constructed by a human programmer but rather learned from test data. Test quality is therefore assessed by neuron and layer-level coverage.

Other researchers have adapted software quality models to define test artifact quality. Neukirchen et al.~\cite{neukirchen_approach_2008} introduced a quality model for TTCN-3 test specifications. They adapted ISO/IEC 9126 to include testing specific characteristics that have no correspondence in the ISO/IEC 9126, such as fault revealing capability, test repeatability, and reusability. Other characteristics were re-interpreted for the context of test code or completely removed from their adaptation of ISO/IEC 9126. The model was limited to unit testing, and focused on metrics for \textit{maintainability} only. The reuse of software quality models adds new dimensions of test quality that have not been covered by Zhu et al.'s three categories of test adequacy: usability (understandability, learnability, operability, test evaluability), maintainability (analyzability, changeability, stability), and reusability (coupling, flexibility, comprehensibility). A similar approach, Athanasiou et al..~\cite{athanasiou_test_2014} introduced a test-code quality model which contained three quality attributes, \textit{test completeness}, \textit{effectiveness} and \textit{maintainability}, and associated metrics. The quality attribute \textit{maintainability} was indirectly based on ISO/IEC 9126. Their analysis of open source code repositories indicates a positive correlation between test code quality and issue handling performance.

Recent research has investigated how engineers' perception of test code can generate useful quality models. Bowes et al.~\cite{bowes_how_2017} conducted a workshop with industry practitioners and elicited 15 unit testing principles and best practices that are expected to result in high quality tests. Tran et al.~\cite{huynh_khanh_test-case_2019} interviewed six practitioners and identified 11 quality characteristics for natural language tests, of which test understandability, simplicity and test step cohesion were mentioned most frequently. Finally, Grano et al.~\cite{grano2020pinsa} interviewed five testing experts and developed a unit test quality taxonomy that includes behavioral, structural and executional facets. They also validated and extended the taxonomy by conducting a survey and study the correlation of existing test quality metrics with the perceived quality collected in the survey~\cite{grano2020pinsa}.

The quality of tests and test suites has also been addressed in research on test smells.
There are different definitions of test smells in the literature.
The concept of test smells was first introduced by Van Deursen et al.~\cite{van2001refactoring} in the context of unit testing for extreme programming (XP).
According to the authors, test smells were derived from code smells and indicate trouble in test code.
They also described 11 types of test smells such as Mystery Guest, Resource Optimism, Test Run War, etc. 
The related test code quality attributes are maintainability, understandability, readability.
Meszaros~\cite{meszaros_xunit_2007} described test smells in the context of using the unit testing framework xUnit.
The author discussed 18 frequent test smells and provided guidelines for analysing causes and eliminating them.
Van Rompaey et al.~\cite{van_detection_2007} discussed test smells with respect to the ideal test execution structure, which includes setup-stimulate-verify-teardown (S-S-V-T cycle).
They defined test smells as "violations of a clean S-S-V-T cycle" and focused on two test smells, General Fixture and Eager Test.
Also, according to Van Rompaey et al., these test smells might reduce specific test quality attributes, including automation, traceability, performance, maintainability, isolation, conciseness, transparency, and explicitness.
The latest secondary study on test smells was conducted by Garousi et al.~\cite{garousi_smells_2018}.
The study provided a catalogue of 139 test smells, a summary of approaches and tools to deal with test smells.
According to the authors, a test smell is "any symptom in test code that indicates a deeper problem".
If they are fixed on time, test smells could lead to several issues, including a decrease in test code quality such as maintainability, readability, fault detection power.

\subsection{Organizing test engineering knowledge}
With the increasing information generated by research comes the need for organizing it to create knowledge. Taxonomies allow to structure knowledge in a way that supports both direct (e.g., stakeholders in a research project) and indirect (e.g., readers of scientific publications) communication~\cite{engstrom_serp-test_2017}. 
Taxonomies provide descriptions of objects and their relationships in a knowledge area~\cite{vegas_maturing_2009,vessey_unified_2005}.
Hence, they offer practitioners and researchers a common understanding of objects and their relationships in the diverse Software Engineering (SE) knowledge field, which eases knowledge sharing and applying findings~\cite{vegas_maturing_2009,vessey_unified_2005}.
Their support in building the knowledge foundation makes it easier to incorporate new knowledge and identify knowledge gaps in SE, especially when the SE knowledge field has been evolving continuously~\cite{vegas_maturing_2009}.
Since there is a wide range of subareas in SE, there is still a need to classify the knowledge in many of those subareas~\cite{usman_taxonomies_2017}.
Therefore, taxonomies have been playing an essential role in maturing the SE knowledge field.

The field of software engineering has seen a surge of taxonomic studies since the 2000's, with 27 focused on software testing alone (1988--2015)~\cite{usman_taxonomies_2017}. Early studies looked at software testing techniques in general~\cite{coward1988review,young1989rethinking}, followed by taxonomies on specific testing aspects, such as fault types~\cite{mariani2003fault,hummer2012deriving,asadollah2015towards}, mutation testing~\cite{yoon1998mutation,becker2012binary,stephan2014towards}, model-based testing~\cite{utting2012taxonomy,felderer2016model}, runtime monitoring~\cite{delgado2004taxonomy,roehm2013monitoring}, security testing~\cite{jiwnani2002maintaining,tian2010research}, unit testing~\cite{vegas_maturing_2009} and regression testing~\cite{ali_search_2019}. The purpose of this paper is to complement this body of knowledge with a test artifact quality model.

\subsection{Tertiary studies}
Secondary studies (systematic literature review and systematic mapping studies) synthesize the knowledge in a particular area by means of a systematic and objective literature collection and analysis process~\cite{kitchenham_evidence_2004}. With the widespread adoption of secondary studies in software engineering, tertiary studies~\cite{kitchenham_systematic_2010} aim at collecting and synthesizing secondary studies\footnote{Quaternary studies have not been observed, yet.}. 

To the best of our knowledge, Garousi et al.~\cite{garousi_systematic_2016} conducted the first tertiary study with the goal of mapping the existing research on software testing. Their review identifies the most frequently studied testing topics (such as model-based testing, regression testing, etc.), the type of stated research questions (a lack of \textit{causality} and \textit{relationship} type), and trends in types, quality and number of primary studies (slow increase in quality and more citations for regular surveys than SLRs and SMs). 
They also report several research areas in software testing which require more secondary studies such as test management, human factors in software testing, exploratory testing and test stopping criteria.
However, their tertiary study did not discuss findings with respect to test-artifact quality, the focus of our tertiary study.
Hence, our contribution is also different as we present the quality model for test artifacts (as discussed in Section~\ref{sec:contribution}) and report relevant context characteristics.
While Garousi et al. used similar and generic keywords for the search process, there was no complete overlap between their selected peer-reviewed secondary studies and ours.
Our first search and second search found 20 and 45 out of 58 peer-reviewed secondary studies selected by Garousi et al. (details on our searches can be found in Section~\ref{sec:searchProcess}).

By snowball sampling starting from Garousi et al.'s review, we identified only one more tertiary study focusing on software testing beyond the one by Garousi et al.\footnote{Garousi et al. cited nine tertiary studies and were cited by thirteen tertiary studies, which in turn were cited by four more tertiary studies, as of October 2020. Only one of those covered software testing (\cite{villalobos-arias_tertiary_2018}).} This tertiary study by Villalobos-Arias et al.~\cite{villalobos-arias_tertiary_2018}, reports a systematic mapping study on areas, tools and challenges in model-based testing (MBT). 
Their main findings are: (a) the two most popular subareas in MBT are UML models and Transition-based notations, and (b) there is a lack of empirical evidence for selecting MBT tools and approaches. They do not report any findings regarding test-artifact quality.

To the best of our knowledge, there is no tertiary study on the quality of test artifacts. Since this topic has been discussed to some extent in secondary studies (e.g.,~\cite{paul_systematic_2014,araki_systematic_2018,shafique_systematic_2013,engstrom_empirical_2008, santos_test_2017}) a synthesis of the state of the art in this area would be helpful for researchers as well as for practitioners.

\subsection{Contribution}\label{sec:contribution}
Test artifact quality has been researched since the dawn of the software engineering discipline. The early review by Zhu et al.~\cite{zhu1997software} focused on unit testing and did not associate test adequacy criteria with a quality model. Such an association would have helped to discover new quality criteria relevant for test artifacts. 

We then identified three later works~\cite{neukirchen_approach_2008, athanasiou_test_2014, grano2020pinsa} which introduced test-code quality models and taxonomies for test artifacts (as explained in Section~\ref{sec:relatedWork_testArtifactQuality}).
While the quality models by Neukirchen et al.~\cite{neukirchen_approach_2008} and Athanasiou et al.~\cite{athanasiou_test_2014} were partly based on the ISO/IEC 9126, the quality taxonomy introduced by Grano et al.~\cite{grano2020pinsa} was based on interviews with practitioners.
In comparison with the two models from Neukirchen et al. and Athanasiou et al., our quality model covers a wider range of quality attributes and quality measurements based on the findings in the literature, and the latest software quality models given by ISO/IEC 25010:2011.

Meanwhile, some quality attributes of our quality model share similar descriptions with some test-quality features of the taxonomy given by Grano et al. even though their names are not related.
For example, the quality feature "(self-)validation" in their taxonomy ("A test should behave as expected,i.e., it must not be defective") is closely related to the quality attributes "reliability" and "consistency" in our model. 
In contrast, their description for "reliability" ("Unit tests should always produce the same results") is not similar with the quality attribute "reliability" in our model. 
There are also quality features mentioned by the practitioners (scope, test design, execution infrastructure) which did not connect to any of our identified quality attributes.
It means that there are still differences between how practitioners describe test-artifact quality and how such quality has been reported in the literature.
Hence, our model does not only provide a more comprehensive picture of how the test-artifact quality was described in the literature, but serves also as a guideline for practitioners to search for knowledge in test-artifact quality in the literature.
In other words, our quality model supports the knowledge transfer between academia and industry.

Various aspects of test engineering knowledge have already been captured with taxonomies. However, the quality of test artifacts has not yet been part of such an effort. Our contribution is to fill this knowledge gap, by means of a tertiary study. This provides insights for researchers on the state-of-art of test artifact quality and supports practitioners in gaining an overview of what aspects of test artifact quality can be applied in which context.
In particular, the test artifact quality model presented in this study can support:
\begin{itemize}
    \item describing new guidelines and templates for designing new test cases.
    \item developing assessment tools for evaluating existing test cases and suites.
\end{itemize}
Furthermore, the model can also be used as a framework for documenting context characteristics to make research results more accessible for research and practice.

\section{Research Method}\label{sec:method}
The goal of this study is to investigate how the quality of test artifacts has been characterized in the context of software engineering in secondary studies.
By test artifact, we refer to documents used for testing, like test cases, test suites, test scripts, test code, test specifications and natural language tests. For this review, we exclude artifacts produced during the execution of tests, such as executions logs, traces or bug reports. 
To achieve this goal, our study aims to answer the following research questions: 
\begin{itemize}
    \item RQ1: What are the differences and similarities among the secondary studies in terms of their characteristics?
    \item RQ2: In which testing-specific contexts have the quality attributes and quality measurements been studied?
    \begin{itemize}
    \item RQ2.1: In which testing-specific contexts has the quality of test artifacts been studied?
    \item RQ2.2: Which quality attributes have been reported for test artifacts?
    \item RQ2.3: Which measurements have been reported to quantify the identified quality attributes?
    \end{itemize}
    \item RQ3: How frequently have quality attributes in particular testing-specific contexts been studied?
    \item RQ4: To which extent is there consensus about the definition of quality attributes and quality measurements?
\end{itemize}

In RQ1, we analyse the characteristics of secondary studies including their quality, review method (systematic literature review, systematic mapping study), and topic.
Comparing the studies' characteristics provides an overview of the field.
For RQ2, we derive the testing-specific contexts from the reviewed secondary studies and which quality aspects have been studied in those contexts.
In particular, we consider two aspects of test artifact quality; quality attributes (RQ2.2) and how they are measured (RQ2.3).
For RQ3, we analyze the data gathered for RQ2 to identify general trends in the study of test artifact quality.
In RQ4, we aim to identify agreements and disagreements regarding definitions of quality attributes and measurements for test artifacts.
This investigation is prompted by the lack of commonly accepted definitions of these aspects of test artifact quality.

\subsection{Search Process}\label{sec:searchProcess}
\begin{table}
{   \footnotesize
    \begin{center}
    \caption{Keywords for search strings}
    \label{tab:keywords}
    \begin{tabular}{p{0.15\linewidth}p{0.35\linewidth}p{0.37\linewidth}}
        \toprule
        \textbf{Sub-string} & \textbf{Keywords in the first search} & \textbf{Keywords in the second search}\\
        \midrule
        Test artifacts & ``test case'' OR ``test suite'' OR ``test script'' OR ``test code'' OR ``test specification'' OR ``natural language test'' &``test case'' OR ``test suite'' OR ``test script'' OR ``test code'' OR ``test specification'' OR ``natural language test'' OR ``test'' OR ``testing'' \\ [4pt]
        Secondary studies & ``systematic review'' OR ``systematic literature review'' OR ``systematic mapping'' OR ``systematic scoping'' & ``systematic review'' OR ``systematic literature review'' OR ``systematic mapping'' OR ``systematic map'' OR ``systematic scoping'' OR ``systematic literature survey'' \\
        \bottomrule
    \end{tabular}
    \end{center}
}
\end{table}

We conducted two rounds of searches to identify secondary studies which discussed the quality of test artifacts.

There are three potential sub-strings for our search strings, namely (1) test artifacts, (2) secondary studies, and (3) quality.
The first sub-string is to find publications related to test artifacts.
The second sub-string is to restrict the searches to the types of secondary studies we are interested in.
We did not add the keyword ``quality'' (the third potential sub-string) to our search string as it would decrease the likelihood of finding relevant publications that did not explicitly use the keyword ``quality'' in their titles and abstracts.
Since ``quality'' is a widely used keyword in many research areas, including it would also lead to an excessive number of irrelevant hits outside the area of software development.
The two selected sub-strings are connected by a Boolean operator \textit{AND} as shown in Table~\ref{tab:keywords}.

The first search, conducted in April 2019, returned 181 publications (121 excluding duplicates) in the following databases and search engines: IEEE Xplore\footnote{\url{http://ieeexplore.ieee.org}}, ACM Digital Library\footnote{\url{http://portal.acm.org}}, ScienceDirect\footnote{\url{https://www.sciencedirect.com/}}, Scopus\footnote{\url{https://www.scopus.com}}, and Google Scholar\footnote{\url{https://scholar.google.com/}} as shown in Table~\ref{tab:searchResults}.

Since the tertiary study by Garousi et al.~\cite{garousi_systematic_2016} is the one closest to our work, we compared our first search's results with their search results to identify potential issues with our search string (considering, however, that we exclude non peer-reviewed, i.e. grey literature).
There was little overlap between the set of publications returned by our first search and Garousi et al.'s initial set of papers.
One issue in our first search was that its keywords for test artifacts did not find potentially relevant publications that used only the general terms ``test'' and/or ``testing'' in their titles, abstracts or keywords.
In our second search, we included the extra keywords ``test'' and ``testing'' for test artifacts. Inspired by Garousi et al.'s search string, we also included further keywords for systematic studies as shown in Table~\ref{tab:keywords}. 

The second search was conducted in October 2019 in Scopus only.
We chose this search engine as it covers most of the major publishers such as IEEE, Elsevier, Springer and ACM.
Also, its high recall (more details in Section~\ref{sec:searchValidation}) showed that the search was reasonably sufficient for our tertiary study.
The second search was restricted to one subject area, ``Computer Science'', to reduce the search noise, and returned 572 publications (569 excluding duplicates) as shown in Table~\ref{tab:searchResults}.
We did not restrict our searches to any specific venues to get as good as possible coverage of peer-reviewed publication venues.
More information regarding the publication venues of the selected secondary studies could be found in Appendix~\ref{appendix:publicationVenues}.

\begin{table}
{    
    \footnotesize
    \begin{center}
    \caption{Search results for first and second search}
    \label{tab:searchResults}
    \begin{tabular}{p{0.09\linewidth}p{0.25\linewidth}p{0.08\linewidth}p{0.4\linewidth}} 
        \toprule
        \textbf{Search} & \textbf{Database/ \mbox{Search Engine}} & \textbf{\# of \mbox{papers}} & \textbf{Search Level}\\
        \midrule
        \multirow{7}{*}{1st} & Scopus & 100 & Title, abstract, keywords \\
        & Google Scholar & 27 & Title \\
        & IEEE & 16 & Title, abstract, keywords \\
        & Science Direct & 23 & Title, abstract, keywords \\
        & ACM & 15 & Title, abstract \\
        \cmidrule{2-3}
        & Total & 181 & \\
        & Excl. duplicates & 121 & \\ [4pt]

        \multirow{2}{*}{2nd} & Scopus & 572 & Title, abstract, keywords \\
        & Excl. duplicates & 569 & \\
        \bottomrule
    \end{tabular}
    \end{center}
}
\end{table}

\begin{table}
{    
    \footnotesize
    \begin{center}
    \caption{Recall and precision of 1st and 2nd searches based on QGS of 13 papers}
    \label{tab:recall_precision}
    \begin{tabular}{l l l}
        \toprule
        & 1st search & 2nd search \\
        Recall & 61.54 & 92.31 \\
        Precision & 6.61 & 2.11 \\
        \bottomrule
    \end{tabular}
    \end{center}
}
\end{table}

\subsection{Search Validation}\label{sec:searchValidation}
To validate our searches, we used a quasi-gold standard (QGS) approach, as suggested by Kitchenham and Charters~\cite{kitchenham2015evidence}.
The concept of using QGS to evaluate search performance in a SLR was first introduced by Zhang et al.~\cite{zhang_identifying_2011}. 
According to Zhang et al., the QGS contains an initial set of papers that are known to be relevant.
Since Garousi et al.'s tertiary study~\cite{garousi_systematic_2016} is the closest related work to our review, we constructed our QGS based on their study's initial set of 121 papers.
Using the same inclusion/exclusion criteria as planned for our study (see Section~\ref{sec:studySelection}), we identified 13 relevant papers from Garousi et al.'s initial set of 121 papers.
Although having more papers in QGS might have been better, there is, to the best of our knowledge, no predefined threshold for the size of QGS suggested by Zhang et al. or other researchers. 
Using a subset of selected papers from another study to assess the automated search's completeness is also a well-known practice supported by other researchers~\cite{kitchenham_systematic_2013}.




We evaluated the performance of the searches by computing recall and precision for each search string.
The recall of a search string is the proportion of known relevant papers found by the search.
The precision is the proportion of the papers found by the search which are relevant to the review.
Table 3 shows the recall and precision of our two searches.
An ideal search would have a high recall and a high precision. 
However, in practice, there is always a trade-off between them in search strategies~\cite{zhang_identifying_2011}. 
In our tertiary study, we aimed for collecting as many relevant secondary studies as possible (a high recall) while accepting more noise in our searches' results (a low precision).
Our approach was aligned with other SLRs in the literature~\cite{zhang_identifying_2011}. 
Hence, despite the low precision of the two searches, their recall is high enough to demonstrate their sufficient performance for using in our tertiary study.



\subsection{Study Selection}
\label{sec:studySelection}
To ensure that we have covered as many relevant papers as possible, we based the study selection process on the union of three sets of papers; the results of our first and second searches (see Section~\ref{sec:searchProcess}) and Garousi et al.'s~\cite{garousi_systematic_2016} initial set of papers.
It is worth to mention that in total, there were 13 papers from Garousi et al.'s study which passed the paper selection criteria.
Those were used as our QGS (see Section~\ref{sec:searchValidation}).
This section describes the whole paper selection process.
However, for the search validation purposes, the selection process was first applied on the 121 papers from Garousi et al.'s study before we selected papers from our own searches.

\begin{figure}
\begin{center}
\includegraphics[width=1\linewidth]{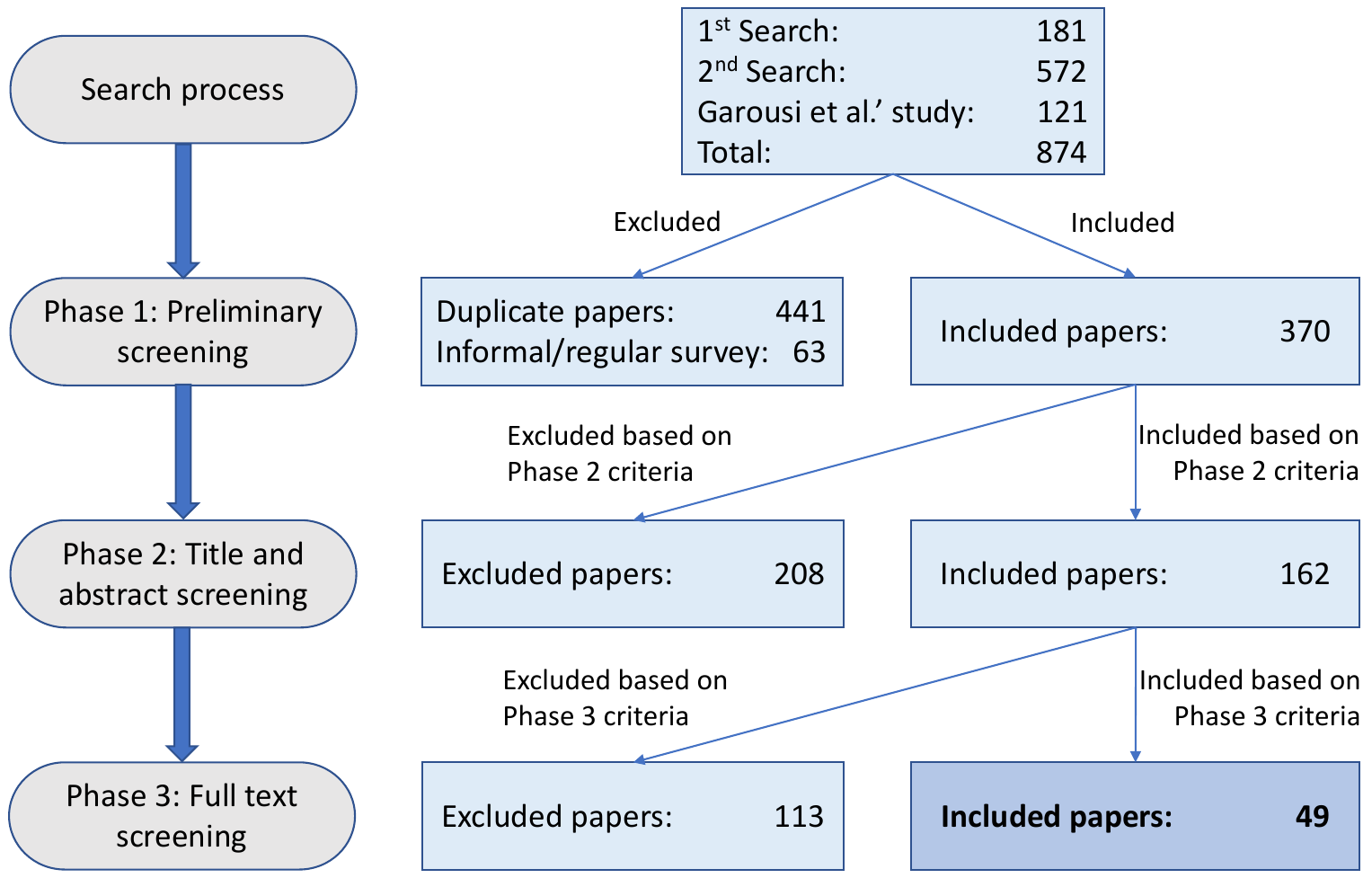}
\caption{Overview of the study selection process}
\label{fig:paperSelectionProcess}
\end{center}
\end{figure}

The study selection process comprised three phases as illustrated in Figure~\ref{fig:paperSelectionProcess} and described below:

\setlist{nolistsep}
\begin{enumerate}[noitemsep]
\item Phase 1: applied on authors, title and abstract
\begin{itemize}
    \item Exclude papers that: 
    \begin{enumerate}[label=(E\arabic*), noitemsep]
        \item are duplicate papers;
        \item are not systematic studies\footnote{Garousi et al.'s~\cite{garousi_systematic_2016} initial set of 121 papers contained 63 informal surveys without research questions. Since we only were targeting systematic studies, these were excluded.};
        \item are not peer reviewed;
        \item are outside computer science or software engineering.
    \end{enumerate}
\end{itemize}

\item Phase 2: applied on title and abstract
\begin{itemize}
    \item Exclude papers that: 
    \begin{enumerate}[label=(E\arabic*), noitemsep]
        \setcounter{enumii}{4}
        \item are not about software testing.
    \end{enumerate}
    \item Include papers that fulfil all of the following:
    \begin{enumerate}[label=(I\arabic*), noitemsep]
        \item are systematic literature reviews (SLR), quasi-SLRs, Multi-vocal literature reviews, or systematic mappings;
        \item discussed or potentially discussed quality of test artifacts
    \end{enumerate}
\end{itemize}

\item Phase 3: applied on full text
\begin{itemize}
    \item Exclude studies that:
    \begin{enumerate}[label=(E\arabic*), noitemsep]
        \setcounter{enumii}{5}
        \item Are duplicate studies (two different studies using the same data)
    \end{enumerate} 
    \item Include studies which discussed any of the following:
    \begin{enumerate}[label=(I\arabic*), noitemsep]
        \setcounter{enumii}{2}
        \item definition of the quality of test artifacts;
        \item quality characteristics of test artifacts;
        \item quality attributes of test artifacts;
        \item quality metrics of test artifacts;
        \item tools, methods, approaches, frameworks to assess test artifacts' quality;
        \item guidelines, checklists to write test artifacts.
    \end{enumerate}
\end{itemize}
\end{enumerate}

\subsubsection{Phase 1: Preliminary screening}
Before applying topic specific selection criteria, the first author removed papers that were duplicates or irrelevant to software engineering or computer science by screening the title, author(s) and abstract.
A duplicate study (E1) in this phase is one with the same author(s), title and abstract as another found study.
Duplicates occurred typically when the same study was in the result set from two or more search engines/databases.
Regarding E4, if a paper's title and abstract was clearly about other fields such as medicine, physics, etc. then the paper was be excluded; otherwise, we included the paper for later assessment by reading the full paper.
After this step, the total number of remaining papers was 370.

\subsubsection{Phase 2: Title and abstract screening}
Before screening the 370 papers, we conducted a pilot study on the criteria to establish a common understanding of the inclusion/exclusion criteria.
Disagreements were discussed and resolved at a face-to-face meeting among the authors.
After that, each of the 370 papers were assessed by two reviewers each.
For each paper, a reviewer first screened its title and abstract against the exclusion criteria to exclude papers which are outside the scope of this review.
After that, the reviewers referred to the inclusion criteria to decide whether that paper should be included for the third phase.
In case a decision was unclear, that paper should be included for the third phase.
This policy helped us to reduce the risk of excluding potentially relevant papers prematurely, i.e. before considering the full text.

\subsubsection{Phase 3: Full text screening}
The selection strategy in the third phase was the same as in the second phase.
Each paper was assessed by two reviewers each.
If a paper was not excluded based on the exclusion criteria, the inclusion criteria were used to evaluate the paper's relevance.
However, instead of reading the titles and abstracts like in the second phase, the reviewers read its full text to make the selection decision.
As a result of the study selection process, we were left with 49 studies for the data extraction process.

\subsubsection{Post-hoc validation}
We calculated the Kappa coefficient to evaluate the extent of reviewers' agreement in study selection.
The Kappa coefficient for each pair of reviewers varied from 0.18 to 0.62, which can be interpreted as a \textit{slight agreement} to \textit{substantial agreement}~\cite{landis_measurement_1977}.
Due to this high variation in disagreement, we reviewed the decisions in face-to-face meetings.
The meetings showed that most disagreements related to one author being slightly more inclusive than the others. All disagreements were resolved in those meetings.

\subsection{Data Extraction}\label{sec:dataExtraction}
We extracted the following data items from the 49 selected secondary studies:
\setlist{nolistsep}
\begin{enumerate}[noitemsep]
    \item Meta-data (Data item 1): publication title, abstract, authors, publication venue, publication year.
    \item Background (Data item 2): information about the study's background, which includes the followings aspects.
    \begin{itemize}
    \item Review method: the adopted literature review method such as systematic literature review, systematic mapping;
    \item Topic of the review: the main focus of the secondary study such as quality of test artifacts, test smells, specific testing techniques or activities.
    \end{itemize}
    \item Quality of test artifacts (Data item 3): information about how the quality of test artifacts was reported.
    \begin{itemize}
        \item Quality attributes: such as re-usability, understandability, etc.;
        \item Quality measurements: such as lines of test code, etc.;
        \item Software testing-specific context: such as testing level, type of software system under test, testing automation level, etc.
    \end{itemize}
\end{enumerate}



To extract the data related to software testing-specific contexts (under data item 3) systematically, we first collected a list of context dimensions that might be considered in a discussion of the quality of test artifacts.
To do so, we performed snowball sampling on the SLR by Tebes et al.~\cite{tebes_systematic_2019}, one of the latest reviews on software testing ontologies and identified 18 sources from the literature and from industry providing taxonomies or ontologies on software engineering~\cite{sapna_ontology_2011,freitas_ontology_2014,vasanthapriyan_ontology-based_2017,vasanthapriyan_ontology-based_2017-1, zhu_developing_2005, rothermel_framework_1994, bourque_guide_2014, ISO, vegas_maturing_2009, ali_search_2019, barbosa_ontology-based_2008, de_campos_regression_2017, de_souza_roost_2017, arnicans_semi-automatic_2013, engstrom_serp-test_2017, usman_taxonomies_2017, cai_test_2009, istqb2015standard}.
From these sources, we selected 14 potential context dimensions such as \textit{Test artifact}, \textit{Test level}, \textit{Testing objective} based on our knowledge/experience and seven relevant sources~\cite{de_souza_roost_2017, vasanthapriyan_ontology-based_2017, bourque_guide_2014, ISO, vegas_maturing_2009, ali_search_2019, usman_taxonomies_2017}. 
Some of these context dimensions contain sub-dimension(s), which we call context factor(s) in our review.
For example, the context dimension \textit{System under test} includes five context factors, namely \textit{Software type}, \textit{Software license}, \textit{Application domain}, \textit{Type of development process} and \textit{Programming language}.
These context dimensions are summarized in Table~\ref{tab:contextDimensions}.

During the data extraction process, we followed the steps described below to extract the context values.
\begin{enumerate}
    \item Step 1: If the authors of the secondary studies explicitly stated that a context value is under a particular context dimension (by mentioning, for example, \textit{system testing} is \textit{testing level}), then we included that context value (\textit{system testing}) to the associated context dimension (\textit{testing level}).
    \item Step 2: In case we identify a context value in a secondary study but the authors of the secondary study did not explicitly state what context dimension the context value belongs to, then we refer to our context dimensions' definitions to make a suitable association.
    For instance, we often identified different context values in the secondary studies such as \textit{test case generation}, \textit{test case design}, \textit{test case prioritisation}, etc. Those context values were not explicitly connected to any context dimensions by the authors of the secondary studies.
    Based on our definitions of context dimensions, we linked those context values to the context dimension \textit{testing activity}.
\end{enumerate} 
The extraction of the context values was done in iterations.
To do so, the first author extracted the values from the secondary studies then discussed them with the other authors to validate the association between the values and their dimensions.

\subsubsection{Data extraction validation}
We followed the following steps to make sure that the data extraction was conducted adequately:

\begin{itemize}
    \item We developed and improved the data extraction form in iterations based on discussion among all authors.
    Hence, the form and the data items were validated.
    \item Data items 1 and 2 (meta-data and background information) were extracted by all authors.
    The extracted information was compared and validated in meetings among all authors.
    \item Regarding data item 3 (test-artifact quality information), data extraction was already considered in the paper selection phase.
    For each relevant paper that contained information regarding definition/assessment of test-artifact quality, the authors added a note about where in the paper this information was found. 
    These notes were reviewed in the validation of the paper inclusion/exclusion step to ensure that they were complete and agreed upon by all authors.
    During the data extraction process, the notes information was used for extracting data regarding the quality of test artifacts. 
    \item We performed a post-hoc validation for data item 3 on four papers (S33, S34, S46, S47) which were selected randomly from the 49 secondary papers.
    Two of those papers were assigned to the second author while the other two papers were assigned to the fourth author. 
    The two authors extracted data independently.
    After that, each of them had a face-to-face meeting with the first author to compare the outcomes and solve disagreements.
    Since there was a high degree of agreement in the extracted data among the authors, we decided not to pursue further validation of the extraction.
    Hence, we decided to use the data extracted by the first author for the data analysis.
\end{itemize}

\subsection{Quality Assessment}
\begin{table*}
{    \footnotesize
    \begin{center}
    \caption{Quality assessment criteria}
    \label{tab:quality_assessment_criteria}
    \begin{threeparttable}
    \begin{tabular}{p{0.03\textwidth}p{0.33\textwidth}p{0.05\textwidth}p{0.45\textwidth}}
        \toprule
        \textbf{ID} & \textbf{Criterion} & \textbf{Score\tnote{1}} & \textbf{Interpretation} \\
        \midrule
        C1 & ``Were inclusion/exclusion criteria reported?'' & Yes & The criteria for paper selection are clearly defined. \\
        & & Partly & The criteria for paper selection are implicit. \\
        & & No & The criteria for paper selection are not defined and cannot be readily inferred. \\
        
        C2 & ``Was the search adequate?'' & Yes & The authors have either (searched for at least four digital libraries AND included additional search strategies) OR (identified and referenced all journals addressing the topic of interest). \\
        & & Partly & The authors have searched three or four digital libraries with no extra search strategies OR they searched a defined but restricted set of journals and conference proceedings. \\
        & & No & The authors searched up to two digital libraries or an extremely restricted set of journals. \\
        
        C3 & ``Were the included studies synthesized?'' & & \\
        C3a & ``Was the evidence actually synthesized and aggregated, or merely summarized?'' & Yes & The synthesis pools the studies in a meaningful and appropriate way. Differences between
        studies are addressed. \\ 
        & & No & There is no real synthesis. The evidences from the individual studies are basically repeated/summarized. \\
        C3b & ``Was the quality of individual studies taken into account in the synthesis?'' & Yes & Yes, to some extent. \\
        & & No & No. \\
        
        C4 & ``Was the quality of the included studies assessed?'' & Yes & The authors have explicitly defined quality criteria and extracted them from each primary study.\\
        & & Partly & The research question involved quality issues that are addressed by the study.\\
        && No & There is no explicit quality assessment of individual papers OR the quality assessment has not been described sufficiently. \\
        
        C5 & ``Are sufficient details about the individual included studies presented?'' & Yes & Information is presented about each paper, so that the data summaries can clearly be traced
        to relevant papers. \\
        & & Partly & Only summary information is presented about individual papers.\\
        & & No & The results for individual studies are not specified. \\
        \bottomrule
    \end{tabular}
    \begin{tablenotes}
    \footnotesize
    \noindent
    \begin{minipage}[c]{1\linewidth}
    \item[1] Yes = 1; Partly = 0.5; No = 0.
    \end{minipage} 
    \end{tablenotes}
    \end{threeparttable}
    \end{center}
}
\end{table*}

We adopted the quality assessment criteria provided by the York University, Centre for
Reviews and Dissemination (CDR) guide for reviews~\cite{dissemination2009systematic} to evaluate the 49 selected secondary studies. These criteria are often used to assess the quality of systematic literature reviews in software engineering~\cite{ali_critical_2019}. The quality assessment criteria are summarized in Table~\ref{tab:quality_assessment_criteria}.

The quality assessment was performed in parallel with the data extraction process.
The first author performed the assessment on all 49 selected studies.
As a post-hoc validation, the second author performed an independent assessment on ten randomly selected studies.
After that, the two authors had a face-to-face meeting to compare and discuss the assessment outcomes.
We found only some minor differences between the two assessments' outcomes.
Once the differences were resolved, the first author adjusted her quality assessment.
Since there was a high degree of agreement in the quality assessment among the authors, we decided not to pursue further validation of the quality assessment.
Consequently, we agreed to use the quality assessment on the rest of the selected studies by the first author.
We did not exclude studies based on their quality score, preserving all available information regarding the quality of test artifacts. 

It is also worth to emphasize that the number of papers chosen for the post-hoc validation of the data extraction and of the quality assessment (four papers and ten papers respectively) are different based on our estimation of the time and effort required to perform the post-hoc validation.
In our tertiary study, extracting data needed more time and effort than assessing the papers' quality.
Therefore, we initially picked more papers for validating the data extraction than the quality assessment.

\subsection{Validity Threats}
\label{sec:validity}
In the following, we discuss potential threats to the validity of this study.

\paragraph{Missing relevant secondary studies.}
One possible threat to the validity of this study is that relevant secondary studies have not been found.
We mitigated this risk by careful development and evaluation of our search strings.
We conducted a second search after comparing our initial set of papers with the papers identified in the related tertiary study on testing by Garousi et al.~\cite{garousi_systematic_2016} to ensure that all relevant papers have been found.
We also used Garousi et al.'s search results as basis for defining a quasi-gold standard~\cite{kitchenham2015evidence} for validating our search results.
The obtained high recall indicates that our search has been sufficiently comprehensive.

\paragraph{Exclusion of grey literature.}
A related validity threat is caused by our decision to exclude grey literature from this study. However, since we review secondary and not primary studies, the risk of excluding relevant but not peer reviewed material is low.

\paragraph{Relevant information from primary studies.}
Since we conduct a tertiary study, our data extraction is based on the information aggregated in the secondary studies. It is therefore possible that relevant information about test artifact quality that was present in primary studies is no longer available in the secondary study.
This could be either because that information has been omitted from the secondary studies or because there is no secondary study available yet covering the missed primary studies.
This threat is inherent to any tertiary study and we accept it here to base our analysis on more widely accepted information.

\paragraph{Bias in paper selection.}
A common threat to the validity of secondary and tertiary studies is a possible bias when the papers to be included in the study are selected.
We mitigate this risk by ensuring that every paper is reviewed by two reviewers.
In this way, the selection does not rely only on the subjective opinion of one reviewer, but on a consensus between the two reviewers instead.
We assessed this consensus with help of the Kappa coefficient between the pairs of reviewers and found slight to substantial agreement.
Furthermore, we conducted face-to-face meetings to resolve any disagreements.

\paragraph{Bias in data extraction.}
Similar to paper selection bias, bias in data extraction is a potential threat to the validity of this study.
We mitigate the risk of data extraction bias by comparing the data extracted by the first author with the data extracted by two of the co-authors for four papers.
In this way, the data extraction approach taken by the first author is validated.
Based on face-to-face discussions, we have found a high degree of consistency between the data extracted by the different reviewers.

\paragraph{Publication bias.}
Publication bias is another common threat to the validity of secondary and tertiary studies.
It refers to the tendency that the studies with negative results are less likely to be published.
We consider that publication bias poses a low risk to the validity of this study, since the failed characterization of test artifact quality does not impact the results of this tertiary study.

\section{Results and Analysis}
\label{sec:results-analysis}
\begin{figure}
\begin{center}

\includegraphics[width=1\linewidth]{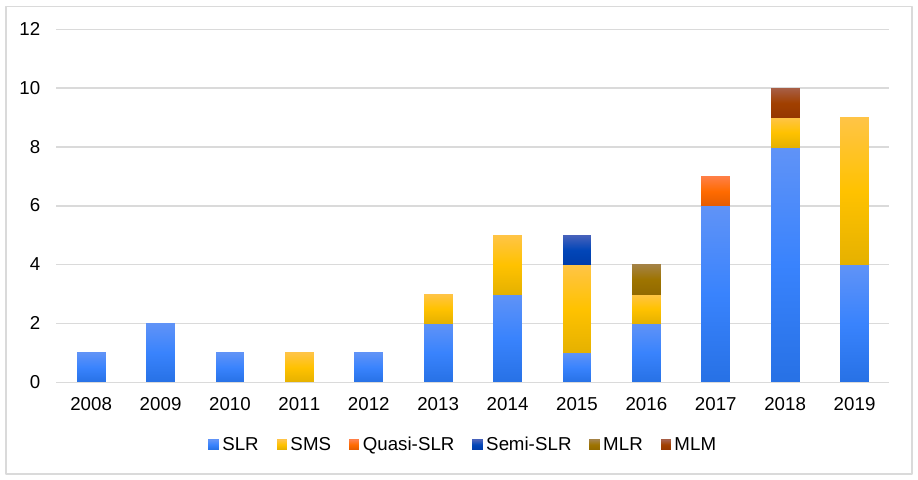}

\caption{Study type distribution over years}
\label{fig:studyTypesOverYears}
\end{center}
\end{figure}

\begin{table*}
{    
    \footnotesize
    \begin{center}
    \caption{List of selected secondary studies}
    \label{tab:selected_studies_1}
    \begin{tabular}{p{0.01\textwidth}p{0.38\textwidth}p{0.15\textwidth}p{0.025\textwidth}p{0.27\textwidth}}
	\toprule
	ID	&	Title	&	Authors	&	Study Type	&	General topic \\
	\midrule
	S1	&	A Comprehensive Investigation of Modern Test Suite Optimization Trends, Tools and Techniques	&	\citeauthor{kiran_comprehensive_2019}	&	SLR	&	Testing activities in regression testing	\\
	S2	&	A systematic literature review of techniques and metrics to reduce the cost of mutation testing	&	\citeauthor{pizzoleto_systematic_2019}	&	SLR	&	Testing techniques	\\
	S3	&	A systematic literature review of test breakage prevention and repair techniques	&	\citeauthor{imtiaz_systematic_2019}	&	SLR	&	Testing activities in regression testing	\\
	S4	&	A Systematic Literature Review of Test Case Prioritization Using Genetic Algorithms	&	\citeauthor{bajaj_systematic_2019}	&	SLR	&	Testing activities in regression testing	\\
	S5	&	A systematic literature review of the test case prioritization technique for sequence of events	&	\citeauthor{ahmad_systematic_2017}	&	SLR	&	Testing activities in regression testing	\\
	S6	&	A systematic literature review on modified condition and decision coverage	&	\citeauthor{paul_systematic_2014}	&	SLR	&	Test data criteria	\\
	S7	&	A systematic mapping addressing Hyper-Heuristics within Search-based Software Testing	&	\citeauthor{balera_systematic_2019}	&	SMS	&	Testing techniques	\\
	S8	&	A systematic mapping study of software product lines testing	&	\citeauthor{da_mota_silveira_neto_systematic_2011}	&	SMS	&	Testing process in specific domain	\\
	S9	&	A systematic mapping study on higher order mutation testing	&	\citeauthor{prado_lima_systematic_2019}	&	SMS	&	Testing techniques	\\
	S10	&	A Systematic Mapping Study on Test Generation from Input/Output Transition Systems	&	\citeauthor{paiva_systematic_2015}	&	SMS	&	Testing techniques	\\
	S11	&	A systematic review of concolic testing with aplication of test criteria	&	\citeauthor{araki_systematic_2018}	&	SLR	&	Testing techniques	\\
	S12	&	A systematic review of search-based testing for non-functional system properties	&	\citeauthor{afzal_systematic_2009}	&	SLR	&	Testing techniques	\\
	S13	&	A systematic review of state-based test tools	&	\citeauthor{shafique_systematic_2013}	&	SLR	&	Testing techniques	\\
	S14	&	A systematic review of the application and empirical investigation of search-based test case generation	&	\citeauthor{ali_systematic_2010}	&	SLR	&	Testing techniques	\\
	S15	&	A systematic review on search based mutation testing	&	\citeauthor{silva_systematic_2017}	&	SLR	&	Testing techniques	\\
	S16	&	A Systematic Review on Test Suite Reduction: Approaches, Experiment's Quality Evaluation, and Guidelines	&	\citeauthor{rehman_khan_systematic_2018}	&	SLR	&	Testing activities in regression testing	\\
	S17	&	Analyzing an automotive testing process with evidence-based software engineering	&	\citeauthor{kasoju_analyzing_2013}	&	SLR	&	Testing process in specific domain	\\
	S18	&	Combinatorial interaction testing of software product lines: A mapping study	&	\citeauthor{sahid_combinatorial_2016}	&	SMS	&	Testing techniques	\\
	S19	&	Continuous Integration, Delivery and Deployment: A Systematic Review on Approaches, Tools, Challenges and Practices	&	\citeauthor{shahin_continuous_2017}	&	SLR	&	Software testing in different types of development process models	\\
	S20	&	Continuous testing and solutions for testing problems in continuous delivery: A systematic literature review	&	\citeauthor{mascheroni_continuous_2018}	&	SLR	&	Software testing in different types of development process models	\\
	S21	&	Effective regression test case selection: A systematic literature review	&	\citeauthor{kazmi_effective_2017}	&	SLR	&	Testing activities in regression testing	\\
	S22	&	Empirical evaluations of regression test selection techniques: A systematic review	&	\citeauthor{engstrom_empirical_2008}	&	SLR	&	Testing activities in regression testing	\\
	S23	&	Factor determination in prioritizing test cases for event sequences: A systematic literature review	&	\citeauthor{ahmad_factor_2018}	&	SLR	&	Testing activities in regression testing	\\
	S24	&	Literature Review of Empirical Research Studies within the Domain of Acceptance Testing	&	\citeauthor{weiss_literature_2016}	&	SLR	&	Software testing in different types of development process models	\\
	S25	&	Machine learning applied to software testing: A systematic mapping study	&	\citeauthor{durelli_machine_2019}	&	SMS	&	Testing automation	\\
	S26	&	Model-based testing for software safety: a systematic mapping study	&	\citeauthor{gurbuz_model-based_2018}	&	SMS	&	Testing techniques	\\
	S27	&	Model-based testing using UML activity diagrams: A systematic mapping study	&	\citeauthor{ahmad_model-based_2019}	&	SMS	&	Testing techniques	\\
	S28	&	Model-driven architecture based testing: A systematic literature review	&	\citeauthor{uzun_model-driven_2018}	&	SLR	&	Testing techniques	\\
	S29	&	On rapid releases and software testing: a case study and a semi-systematic literature review	&	\citeauthor{mantyla_rapid_2015}	&	semi-SLR	&	Software testing in different types of development process models	\\
	S30	&	Problems, causes and solutions when adopting continuous delivery—A systematic literature review	&	\citeauthor{laukkanen_problems_2017}	&	SLR	&	Software testing in different types of development process models	\\
	S31	&	Quality Factors of Test Cases: A Systematic Literature Review	&	\citeauthor{barraood_quality_2018}	&	SLR	&	Test case quality	\\
	S32	&	Regression testing of web service: A systematic mapping study	&	\citeauthor{qiu_regression_2014}	&	SMS	&	Testing activities in regression testing	\\
	\midrule
	\multicolumn{5}{l}{Continued in Table~\ref{tab:selected_studies_2}} \\
	\bottomrule
    \end{tabular}
    \end{center}
}
\end{table*}

\begin{table*}
{    
    \footnotesize
    \begin{center}
    \caption{List of selected secondary studies (continued)}
    \label{tab:selected_studies_2}
    \begin{tabular}{p{0.01\textwidth}p{0.38\textwidth}p{0.15\textwidth}p{0.025\textwidth}p{0.27\textwidth}}
	\toprule
	ID	&	Title	&	Authors	&	Study Type	&	General topic \\
	\midrule
	S33	&	Smells in software test code: A survey of knowledge in industry and academia	&	\citeauthor{garousi_smells_2018}	&	MLM
	&	Test code and relevant smells	\\
	S34	&	Software test-code engineering: A systematic mapping	&	\citeauthor{garousi_yusifolu_software_2015}	&	SMS	&	Test code and relevant smells	\\
	S35	&	Software testing with an operational profile: OP definition	&	\citeauthor{smidts_software_2014}	&	SMS	&	Testing techniques	\\
	S36	&	Specifications for Web Services Testing: A Systematic Review	&	\citeauthor{nabil_specifications_2015}	&	SLR	&	Testing activities in specific domain	\\
	S37	&	Systematic literature review on regression test prioritization techniques	&	\citeauthor{singh_systematic_2012}	&	SLR	&	Testing activities in regression testing	\\
	S38	&	Test case design for context-aware applications: Are we there yet?	&	\citeauthor{santos_test_2017}	&	Quasi-SLR	&	Testing activities in specific domain	\\
	S39	&	Test case prioritization approaches in regression testing: A systematic literature review	&	\citeauthor{khatibsyarbini_test_2018}	&	SLR	&	Testing activities in regression testing	\\
	S40	&	Test case prioritization: A systematic mapping study	&	\citeauthor{catal_test_2013}	&	SMS	&	Testing activities in regression testing	\\
	S41	&	Test case prioritization: A systematic review and mapping of the literature	&	\citeauthor{de_campos_junior_test_2017}	&	SLR	&	Testing activities in regression testing	\\
	S42	&	Test case selection: A systematic literature review	&	\citeauthor{narciso_test_2014}	&	SLR	&	Testing activities in regression testing	\\
	S43	&	Testing techniques selection: A systematic mapping study	&	\citeauthor{santos_testing_2019}	&	SMS	&	Testing techniques	\\
	S44	&	The approaches to quantify web application security scanners quality: A review	&	\citeauthor{seng_approaches_2018}	&	SLR	&	Penetration testing	\\
	S45	&	The experimental applications of search-based techniques for model-based testing: Taxonomy and systematic literature review	&	\citeauthor{saeed_experimental_2016}	&	SLR	&	Testing techniques	\\
	S46	&	Unit testing approaches for BPEL: A systematic review	&	\citeauthor{zakaria_unit_2009}	&	SLR	&	Testing techniques	\\
	S47	&	Vertical Test Reuse for Embedded Systems: A Systematic Mapping Study	&	\citeauthor{flemstrom_vertical_2015}	&	SMS	&	Testing activities in regression testing	\\
	S48	&	Web application testing: A systematic literature review	&	\citeauthor{dogan_web_2014}	&	SLR	&	Testing process in specific domain	\\
	S49	&	When and what to automate in software testing? A multi-vocal literature review	&	\citeauthor{garousi_when_2016}	&	MLR
	&	Testing automation	\\
	\bottomrule
    \end{tabular}
    \end{center}
}
\end{table*}

\begin{figure*}
\begin{center}

\includegraphics[width=1\textwidth]{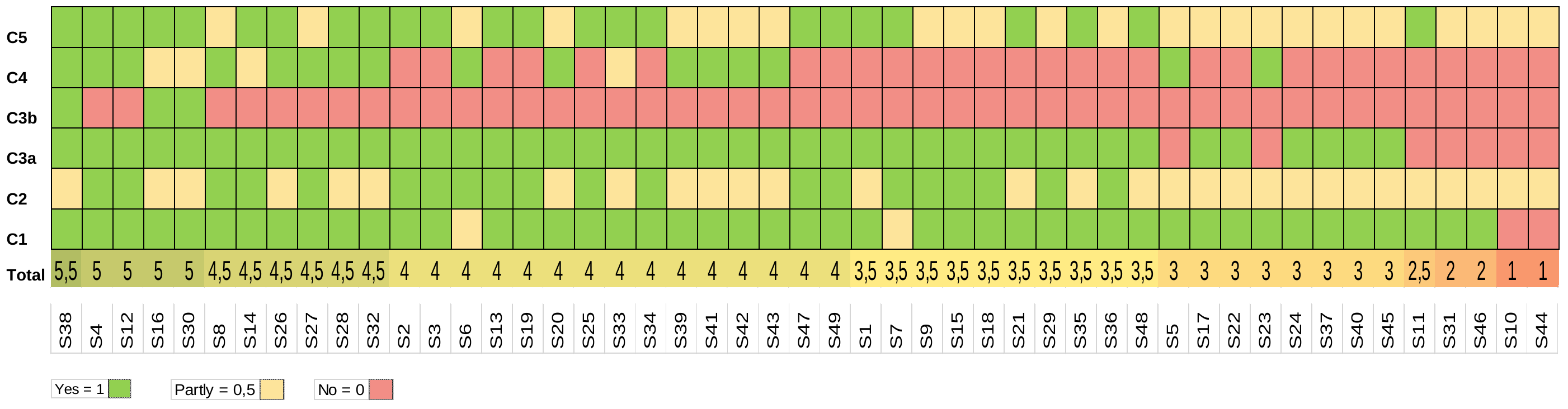}

\caption{Quality Score Overview}
\label{fig:quality_score_sorted}
\end{center}
\end{figure*}

We organize the presentation of results and analysis according to our four research questions.
\subsection{RQ1 -- Differences and similarities between secondary studies}
As described in Section~\ref{sec:studySelection}, we selected 49 secondary studies.
The complete list of the selected studies is summarized in Table~\ref{tab:selected_studies_1} and Table~\ref{tab:selected_studies_2}.
The majority of the studies are Systematic Literature Reviews (SLRs) (31 studies), and Systematic Literature Mappings (SMSs) (14 studies).
There are only two studies described as semi-SLR (S29) and quasi-SLR (S38) as the authors stated that they included studies which were not found by the search strings, and they did not have a meta-analysis respectively.
There are two other studies (S33, S49) which involved non-published and not peer-reviewed sources of information, which is often called \textit{grey literature}, and therefore described as Multi-vocal Literature Review (MLR) and Multi-vocal Literature Mapping (MLM).

The distribution of the studies over years is shown in Figure~\ref{fig:studyTypesOverYears}.
Overall, the number of secondary studies increased gradually from 2008 to 2019.
No study published before 2008 fulfilled our study selection criteria.
As our search was concluded in October 2019, it is reasonable that the number of selected studies published in 2019 is slightly lower than in 2018.
The increasing trend of secondary studies could be highly related to the introduction of evidence-based software engineering by Kitchenham et al.~\cite{kitchenham_evidence_2004}.

We have identified different topics covered in the selected studies. 
The most dominant one is about different testing techniques (17 studies) which include model-based testing (S10, S26, S27, S28, S35, S45), mutation testing (S2, S9, S15), search-based testing (S7, S12, S14), concolic testing (S11), combinatorial interaction testing (S18), different testing techniques in unit testing (S46), and approaches for selection of testing techniques and testing tools (S13, S43).
The second common topic is about testing activities in regression testing (15 studies).
The covered activities are test case selection (S21, S22, S32, S42), test case prioritisation (S4, S5, S23, S32, S37, S39, S40, S41), test case reuse (S47), test case repair (S3), and test suite reduction (S1, S16, S32).
Studies in the third common topic are about software testing in various types of development processes which are continuous practices (S19, S20, S30), rapid releases (S29) and test-driven development (S24).

The less frequent topics are about the general testing process in specific domains (software product lines, automotive systems, web applications) (S8, S17, S48), testing activities in specific domains (web services, context-aware applications) (S36, S38), testing automation (S25, S49), and test code and smells (S33, S34).

Among the least frequent topics, it is worth to note that we found only one SLR conducted by Barraood et al. (S31) which had its main focus on quality of test cases and test suites.
Barraood et al. (S31) identified 30 quality metrics for test cases, three of which did not have descriptions.
However, among the other 27 metrics, if we refer to their provided descriptions, there are six quality attributes and 21 quality metrics for both test cases and test suites.
Our test-artifact quality model (see details in Section~\ref{sec:rq.2.2.quality_attribute}) covers all of their quality attributes that have descriptions.
Barraood et al. also concluded that the quality of test-case metrics influences test-case effectiveness.

Our main concern with this secondary study is its low quality in terms of review method and synthesis.
As demonstrated by our quality assessment (see Table~\ref{tab:quality_assessment_result} in Appendix~\ref{appendix:qualityAssessmentResult}), this secondary study only summarized findings reported in their selected primary studies (C3, C3a, C3b).
Also, there are some inconsistencies in their research method.
While they indicated that their secondary study was about the quality of test cases, they limited their search terms to \textit{effectiveness} of test cases, and included papers presenting "metrics of testing quality" and "good test cases" as mentioned in their inclusion criteria.
They also did not assess the quality (C4) or provide sufficient details (C5) of the selected primary studies.
Therefore, even though the study's focus is very close to our tertiary study's aim, we could not rely completely on its findings and conclusions.
Hence, we still see a knowledge gap that we address in our comprehensive review.

Another vital factor which could demonstrate differences and similarities among the selected secondary studies is their quality scores, as illustrated in Figure~\ref{fig:quality_score_sorted}.
Based on our quality assessment criteria (see Table~\ref{tab:quality_assessment_criteria}), a secondary study could obtain 0 to 6 points. 
Even though there is no study which fulfilled all the assessment criteria, the secondary studies had to some extent adequate search procedures (C2) and provided sufficient details about their selected primary studies (C5).
Most of the secondary studies (47 studies) also reported their inclusion/exclusion criteria for primary studies selection (C1).

According to CRD/DARE, the three quality criteria C1, C2, C3 (including two sub-criteria C3a and C3b) are considered more important than the other two criteria (C4 and C5).
Among our selected secondary studies, there is no study which completely fulfilled all the important quality criteria (C1, C2, C3a, C3b).
In this regard, while the criteria C1 and C2 were often addressed in most of the secondary studies, we found that the way findings were synthesized in the secondary studies was not thorough according to the criteria C3a, C3b.
On one hand, the evidences from primary studies were synthesized in most of the studies (42 studies).
On the other hand, only three of those secondary studies considered the quality assessment results of their primary studies in the synthesis (C3b).

It is worth to note that despite the observation above, nearly half of the secondary studies (20 studies) did define quality assessment criteria for their selected primary studies or have some research question(s) involving quality issues (C4).
Nevertheless, most (17 out of 20 studies) used the quality assessment results as a threshold for their primary study inclusion only.
\begin{table*}
{    
    \footnotesize
    \begin{center}
    \caption{Context Dimensions} 
    \label{tab:contextDimensions}
    \begin{tabular}{p{0.16\textwidth}p{0.165\textwidth}p{0.6\textwidth}} 
        \toprule
        \textbf{Context Dimension} & \textbf{Context Factor} & \textbf{Context Value reported in the selected studies}\\
        \midrule
        Test artifact & Type &  test case, test suite (of test cases, test scripts, test-code scripts)  \\
        \midrule
        Test level &  & integration testing, system testing, unit testing  \\
        \midrule
        Testing objective & & acceptance testing, compatibility testing, execution time testing, penetration testing, quality of service testing, regression testing, robustness testing, safety testing, security testing, UI testing, usability testing  \\
        \midrule
        Testing activity & & test case design, test case execution, test case generation, test case prioritization, test case selection, test coding, test data generation, test script generation, test script repair, test suite reduction \\
        \midrule
        Testing technique & & combinatorial interaction testing, concolic testing, model-based testing (MBT), mutation testing, search-based software testing, state-based testing \\
        \midrule
        Testing approach & & black-box testing, white-box testing \\
        \midrule
        Testing frequency & & continuous \\
        \midrule
        \multirow{3}{*}{System under test} & Software type & academic experimental/simple code examples, context-aware software system, desktop application, distributed business application, embedded system, GUI-based application, mobile application, reactive system, safety-critical application, software product line, web-service-based system \\
        & Software license & commercial, open source \\
        & Application domain & civil avionics \\
        & Type of development process & acceptance test driven development, continuous practices (integration/delivery/deployment), rapid releases \\
        & Programming language & Business Process Execution Language (BPEL), C, C++, JavaScript \\
        \midrule
        Testing tool and framework & &  Add-on for Rational Rose (TDE/UML), AGTCG, Architectural model for branch coverage Enhancement (ABCE), ATOM, ATUSA, AUSTIN, Cadnece SMV model checker, CATG, CESE, COLT, COMEDY, Concolic execution tool, CrashFinderHB, CRAX, CREST, CREST-BV, CUTE, DART, Evacon, GERT, Green Analysis of Branch Coverage Enhancement, GUIAnalyzer, iConSMutate, jCUTE, jDart, jFuzz, JSART, Junit, KLEE, LCT, LLSPLAT, PathCrawler, Plugin for Eclipse Modeling Framework, SAGE, SCORE, SITAR (QTP), SMT Solver, SPLAT, Star (Software Testing by Abstraction Refinement), SynConSMutate, TESTEVOL, Trex, TSGAD, TTCN, Ttworkbench, UMLTGF, UTG, Yices \\
        \midrule
        \multirow{3}{*}{Automation level} & Automation activity	&  test case execution, test case generation, test data generation, test script execution, test script generation, test script repair \\
        & Automation degree & \textit{not discussed in selected studies} (e.g: semi automation, full automation) \\
        \midrule
        \multirow{3}{*}{Orientation} & Technology orientation	&  Yes/No (example cases for ``Yes'': MBT tools, test case design techniques, test suite optimisation techniques, testing techniques selection approaches, etc.)\\
        & People orientation & Yes/No (example cases for ``Yes'': developers, software engineers, testers, etc.) \\
        \midrule
        \multirow{3}{*}{Solutions/Improvements} & Correction & \textit{not discussed in selected studies} \\
        & Detection & \textit{not discussed in selected studies} \\
        & Prevention & \textit{not discussed in selected studies} \\
        \midrule
        Quality standards & & \textit{not discussed in selected studies} (e.g: ISO, ISTQB) \\ 
        \midrule
        Quality type & & \textit{not discussed in selected studies} (e.g: Quality In Use, Product Quality) \\ 
        \bottomrule
    \end{tabular}
    \end{center}
}
\end{table*}

\subsection{RQ2 -- Quality attributes and measurements in testing-specific contexts}
We report here the results that cover testing-specific contexts, quality attributes and quality measurements.

\subsubsection{RQ2.1 -- Testing-specific contexts}\label{sec:rq2.1.context}

With this research question, we focused on reporting and analyzing the testing-specific contexts in which the quality of test artifacts has been discussed in the literature.
In our review, such a context is built by different context dimensions and the values under each dimension as discussed in Section~\ref{sec:dataExtraction}.
As mentioned in the same section, we proposed 14 potential context dimensions.
However, only 11 of the potential context dimensions were found in the selected secondary studies while the other four dimensions, including \textit{Automation degree}, \textit{Solutions/Improvements}, \textit{Quality standards}, and \textit{Quality type}, were not.
The reported context dimensions and their context values found in the selected studies are presented in Table~\ref{tab:contextDimensions}.

How often such a dimension has been used could reveal its relevance in terms of describing the contexts in which the test artifact quality has been studied in general.
In our review, we presented a dimension's frequency by the number of secondary studies in which that dimension was reported.
As shown in Table~\ref{tab:context_frequency}, there is clear gap between the most common context dimensions and the least common ones.

The most common context dimensions which were reported in at least 15 studies are \textit{Test artifact}, \textit{Testing activity}, \textit{Orientation}, \textit{Testing objective}, \textit{Automation activity}, \textit{Testing technique}, and \textit{Software type} of system under test (SUT).
It is worth to note that with the most common context dimension, \textit{Test artifact}, there is only one selected study (S12) which is not associated with it.
In the study, fitness functions were described as test adequacy criteria.
However, there is no information which could help to connect the fitness functions to any value under the \textit{Test artifact} dimension.

In contrast to the most common context dimensions, it appears that other dimensions related to SUT, including \textit{type of development process}, \textit{application domain}, \textit{programming language},\textit{ software license} have not been often reported (in five papers at most) in connection with test artifact quality.
Likewise, dimensions such as \textit{testing level}, \textit{testing approach}, \textit{testing frequency}, \textit{testing tool and framework} have not been often reported when test artifact quality was discussed.

Table~\ref{tab:context_frequency} also shows the frequencies of the most dominant context values found under each context dimension.
Among those dominant context values, \textit{test suite}, \textit{test case}, and \textit{regression testing} are the most frequent of all.
Meanwhile, the context values under the least frequent context dimensions were consequently the least common context values.

\begin{table}
{    
    \footnotesize
    \begin{center}
    \caption{Context Frequency}
    \label{tab:context_frequency}
    \begin{tabular}{ll}
	\toprule
	Context Dimension	&	Frequency	\\
	(> Context Value)   &   (\# of secondary studies) \\
	\midrule
	Test Artifact	&	48	\\
    > test suite	&	34	\\
	> test case	&	20	\\
	\midrule
	Testing activity	&	34	\\
	> test case generation	&	9	\\
	> test case prioritisation	&	8	\\
	\midrule
	Orientation	&	33	\\
	> technology orientation	&	29	\\
	\midrule
	Testing objective	&	20	\\
	> regression testing	&	15	\\
	\midrule
	Automation activity	&	17	\\
	> test case generation	&	9	\\
	\midrule
	Testing technique	&	15	\\
	> MBT	&	7	\\
	\midrule
	SUT\_Software type	&	15	\\
	> safety-critical application	&	3	\\
	> software product line	&	3	\\
	> web-service-based system	&	3	\\
	\midrule
	SUT\_Type of development process	&	5	\\
	> continuous practices	&	2	\\
	\midrule
	Testing tool and framework	&	5	\\
	> ATOM	&	2	\\
	\midrule
	Test level	&	4	\\
	> system testing	&	2	\\
	\midrule
	Testing approach	&	3	\\
	> black-box testing	&	3	\\
	\midrule
	Testing frequency	&	3	\\
	> continuous	&	3	\\
	\midrule
	SUT\_Application domain	&	2	\\
	> each context value (see Table~\ref{tab:contextDimensions})  &	1	\\
	\midrule
	SUT\_Programming language	&	2	\\
	> each context value (see Table~\ref{tab:contextDimensions})  &	1	\\
	\midrule
	SUT\_Software License	&	1	\\
	> each context value (see Table~\ref{tab:contextDimensions})  &	1	\\
	\bottomrule
    \end{tabular}

    \end{center}
}
\end{table}

\subsubsection{RQ2.2 -- Quality Attributes}\label{sec:rq.2.2.quality_attribute}
\begin{figure*}
\begin{center}
\includegraphics[width=1\textwidth]{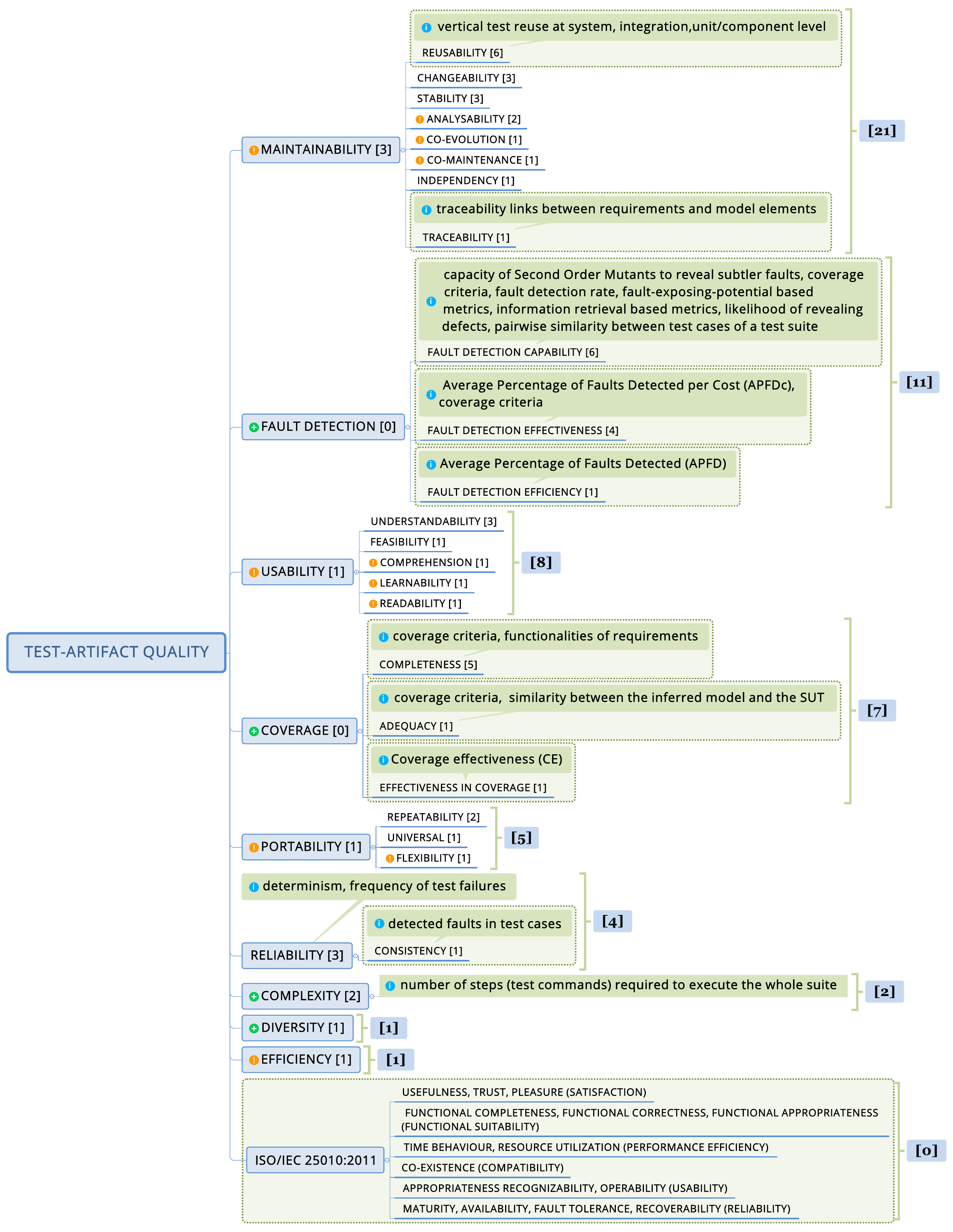}
\caption{Test Artifact Quality Model \\ \protect\includegraphics[height=0.35cm]{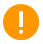} quality attributes (QA) without description;
\protect\includegraphics[height=0.35cm]{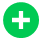} QAs without connection to ISO/IEC 25010:2011; \\ 
\protect\includegraphics[height=0.35cm]{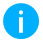} quality measurement(s) associated with the QA;
\textbf{[ ]} next to each QA is the number of unique contexts in which the QA was studied. Each number on the outer right is the total number of unique contexts for the corresponding group of QAs
}
\label{fig:test_artifact_quality_model}
\end{center}
\end{figure*}

\begin{table*}
{    
    \footnotesize
    \begin{center}
    \caption{Test Artifact Quality Attribute (QA) Description \textit{(sorted according to Figure~\ref{fig:test_artifact_quality_model})}}
    \label{tab:qa_description}
   \begin{tabular}{p{0.07\textwidth}p{0.11\textwidth}p{0.6\textwidth}p{0.02\textwidth}}
	\toprule
	Main QA  &   Sub-QA &   Description (described by the authors of the secondary studies)	&	Ref	\\
	
	\midrule
	Maintain-ability &   Reusability &	\textbf{(1) By} vertical test reuse, we refer to reuse of test cases between different levels of integration. For example, test cases designed for component- or unit-level testing can (with or without modifications) be reused at the subsystem integration or system test levels.	&	S47	\\
	&   &	(2) The same test can be reused as a part of another test, e.g., a login test must pass in a web-based application before other tests can be executed.	&	S49	\\
	&   Changeability	&	\textbf{Changeable} structure and style of a test case which allows changes to be made easily, completely, and consistently.	&	S31	\\
	&   Stability	&	\textbf{(1) It} is about stability of the logic under test.	&	S49	\\
	&   &   (2) Unstable test case is likely to break or not reflecting the functionality to be tested. &   S30 \\
	&   Independency	&	\textbf{The} measurement of the degree of dependency among one test case to other test cases.	&	S31	\\
	&   Traceability	&	\textbf{Requirements} traceability refers to the ability to link requirements, often created by a third-party tool (e.g., IBM DOORS), to parts of the test model (e.g., a transition, a path) and, therefore, to test cases. Traceability provides several benefits, such as identifying which test cases exercise which requirements, which requirements are still to be tested and which requirements are linked to a failed test.	&	S13	\\
	\midrule
	
	Fault Detection (FD) &  FD capability	&   \textbf{(1) Higher} code coverage can be a good indicator of fault detection capability. There are a number of coverage criteria for code coverage, such as function coverage, statement coverage, branch coverage,and condition coverage.	&	S23	\\
	&   &	(2) The test’s likelihood of revealing defects;	&	S49	\\
	&   &	(3) The capability of the set cases have to detect defects;	&	S43	\\
	&   &	(4) The ability to identify faults in the source code, code coverage.	&	S7	\\
	&   FD effectiveness	&	\textbf{(1) In} the absence of precise quality indicators for test suites, the coverage of a test suite is usually used as proxy for its fault-detection effectiveness.	&	S25	\\
	&   &	(2) It considers a test case to be effective in the current release if the same test was also able to detect faults in previous releases.	&	S31	\\
	&   FD efficiency	&	\textbf{How} quickly an arranged and optimized test suite can discover defects.	&	S39	\\
	\midrule
	
	Usability   &   Understandability	&	\textbf{How} easy to understand a test case in terms of its internal and external descriptions. Internal information is a test case's content which is actually used in software testing, and it is most cared to test engineer (e.g.: test  target, tested function, test  scenario, test  input, test  step,  test  expected result). External information is interface attributes of test case which may help user understand test case, ease retrieval procedure and improve test case reuse (e.g.: keyword, precursor test case, test field).	&	S31	\\
	&   Feasibility	&	\textbf{In} the context of GUIs, test cases take the form of sequences of events that are executed in hopes of detecting faults in the application. However, test cases might be rendered infeasible if one or more events in the sequence are disabled or inaccessible. This type of test case terminates prematurely and end up wasting resources.	&	S25	\\
	\midrule
	
	Coverage    &   Completeness	&	\textbf{(1) Incomplete} acceptance test cases means that not all functionalities of the requirements are executed;	&	S24	\\
	&   &	(2) The intended mutant coverage that the current test suite need to evolve in order to achieve;	&	S2	\\
	&   &	(3) How completely a test suite exercises the capabilities of a piece of software.	&	S8	\\
	&   Adequacy	&	\textbf{An} adequate test suite is one that implies that the SUT is free of errors if it runs correctly.	&	S25	\\
	&   Effectiveness in coverage	&	\textbf{Code} coverage (CE) is a metric that integrates the size of test suites and the coverage of each test case. CE values range from zero to one, where a higher value indicates a better effectiveness in coverage.	&   S39   \\
	\midrule
	
	Portability &   Repeatability	&	\textbf{The} number of environments to test a SUT in usually increases test repeatability, e.g., when testing Android applications, one needs to repeat the same test in different Android phone models.	&	S49	\\
	&   Universal	&	\textbf{It} is reflected from test scenarios and test fields in which a test case can be executed.	&	S31	\\
	\midrule
	
	Reliability &   Reliability &	\textbf{(1) A} test that fails randomly is not reliable;	&	S20	\\
	            &   &	(2) Unreliable tests means frequent test failures.	&	S19	\\
	&   Consistency	&	\textbf{Inconsistent} acceptance test cases means that faults in the test cases are present, which is typically revealed during implementation.	&	S24	\\
	\midrule
	
	Complexity  &   &   \textbf{Test suite} complexity is defined as the number of steps (test commands) required to execute the whole suite.	&	S48	\\
	
	\midrule
	Diversity   &   &   \textbf{It} is about an overlap of test cases being executed between subsequent releases.	&	S29	\\
	\bottomrule
\end{tabular}

    \end{center}
}
\end{table*}

\begin{table*}
{    
    \footnotesize
    \begin{center}
    \caption{Test Artifact Quality Measurement (QM) Description \textit{(sorted according to Figure~\ref{fig:test_artifact_quality_model})}}
    \label{tab:QA_QMDescription}
    \begin{tabular}{p{0.07\textwidth}p{0.1\textwidth}p{0.15\textwidth}p{0.5\textwidth}p{0.02\textwidth}}
	\toprule
	Main QA &   Sub-QA  &   QM	&	QM Description (as described by the authors of the secondary studies)	& Ref   \\
	\midrule
	Fault detection &   Fault detection capability	&	Capacity of SOM to reveal subtler faults	&	\textbf{Subtler} faults are those which are not revealed by its constituent First Order Mutants (FOMs).	&	S7\\
	&   &	Coverage criteria	&	\textbf{Coverage} of significant service-related elements (Service Activities (SAs) and Service Transitions (STs)), which originated from statements and branches in traditional software artifacts, are the most commonly used criteria	&	S32\\
    &   &	Fault-exposing-potential (FEP) based metrics    &   \textbf{FEP} is the probability of detecting faults for each test case. The metrics include (1) Potency: the probability of a test case to detect a fault based on the predicted test oracles; (2) Change sensitivity: measure the importance of test cases based on the assumption that sensitive test cases that potentially execute more service changes have a higher ability to reach the faults caused by changes; (3) Fault rate: detected seeded faults w.r.t. the execution time; (4) Fault severity: based on the combination of fault rate and fault impact (importance of a fault).	&	S32\\
	&   &	Information-retrieval based metrics	&	\textbf{If} the execution history of one test case covers more identifiers (e.g: method signature, data structure in services) emerged simultaneously in the service change descriptions, the test case has a higher probability to cover the potential faults caused by this change. Based on this hypothesis, the degree of matching between service change query and service execution history can be used to measure the priority of the test cases. The test case with the highest similarity is ranked first.	&	S32\\
    &   Fault detection efficiency	&	Average Percentage of Faults Detected (APFD)	&	\textbf{APFD} calculates the weighted average of the percentage of faults detected by the test cases of the test suite. The result of APFD ranges from zero to 100, where a greater value indicates a better fault revealing rate.	&	S39\\
	\midrule
	Coverage    &   Completeness   &   Coverage criteria & \textbf{The} subcategories include statement coverage, branch coverage, method or function coverage, condition coverage, requirement coverage, and test case coverage	&   S21\\
    &   &   Coverage criteria & \textbf{Two} coverage criteria for framework-based product lines: hook and template coverage, that is, variation points open for customisation in a framework are implemented as hook classes and stable parts as template classes. They are used to measure the coverage of frameworks or other collections of classes in an application by counting the structures or hook method references from them instead of single methods or classes.	&   S8\\
    &   Adequacy    &	Coverage criteria	&	\textbf{Behavioral} coverage is essentially concerned with inferring a model from a system by observing its behavior (i.e., outputs) during the execution of a test suite. If one can show that the model is accurate, it follows that the test suite can be considered adequate.	&   S25\\
	&   Effectiveness in coverage	&	Coverage effectiveness (CE)	&	\textbf{It} is a metric that integrates the size of test suites and the coverage of each test case. It is the ratio between the size of the whole test suite and the coverage of reordered test suite that reveals all faults or meets all requirements. CE values range from zero to one, where a higher value indicates a better effectiveness in coverage.	&	S39\\
	\bottomrule
    \end{tabular}
    \end{center}
}
\end{table*}

In this section, we report the quality attributes of test artifacts associated with the identified test-specific contexts discussed in Section~\ref{sec:rq2.1.context}.
We constructed a quality model, shown in Figure~\ref{fig:test_artifact_quality_model}, to provide an overview of the 30 identified quality attributes.
The quality model was inspired by the software quality models in the ISO/IEC 25010:2011.
Among the 30 quality attributes identified in the secondary studies, 19 had descriptions while the rest were only mentioned by name.
The quality attributes without descriptions are marked with orange exclamation symbols in Figure~\ref{fig:test_artifact_quality_model}.

We categorized the identified quality attributes based on the hierarchy in the quality-in-use model and the product quality model provided in the ISO/IEC 25010:2011.
As a result, there are nine main quality attributes in our model: \textit{Coverage}, \textit{Fault detection}, \textit{Portability}, \textit{Maintainability}, \textit{Reliability}, \textit{Usability}, \textit{Complexity}, \textit{Diversity}, and \textit{Efficiency}.
Among those nine attributes, \textit{Fault detection} and \textit{Coverage} were created by the authors to group other quality attributes which were discussed in the secondary studies, but could not be fitted in an existing category provided by ISO/IEC 25010:2011.
In contrast to the other attributes, \textit{Complexity}, \textit{Diversity}, and \textit{Efficiency} do not have any sub-attributes.

Overall, there are two main differences between our quality model and the quality models defined by the ISO standard.
First, the four quality attributes \textit{Fault detection}, \textit{Coverage}, \textit{Complexity}, and \textit{Diversity} do not have any connection or similarity with the quality characteristics defined in the ISO standard.
This is to be expected, since the ISO standard is defined for software, while these four attributes focus on specific aspects of test artifacts, which are not applicable for software in general.
For example, \textit{Coverage} can be defined for test artifacts to describe the degree to which the functionality of a system under test is tested.
For software in general, this quality attribute is meaningless.
These additional quality attributes are marked with green symbols in Figure~\ref{fig:test_artifact_quality_model}.

Second, we attached a number (in square brackets) to each quality attribute to show the number of unique contexts in which a quality attribute was studied.
As \textit{Fault detection} and \textit{Coverage} were only created to group the identified quality attributes, the number of unique contexts in which those two attributes were discussed is zero.
\textit{Maintainability} and its sub-attributes are the most frequent quality attributes with 21 unique contexts in total, followed by the sub-attributes under \textit{Fault detection} with 11 unique contexts. 
\textit{Usability} and its sub-attributes were also discussed widely in the secondary studies with eight unique contexts in total.
Meanwhile, \textit{Complexity}, \textit{Diversity}, and \textit{Efficiency} are the least frequent quality attributes with at most two unique contexts each.

To provide a broader picture of test artifact quality attributes, we also included in Figure~\ref{fig:test_artifact_quality_model} a set of 15 quality attributes from the ISO/IEC 25010:2011 which could be relevant to test artifact quality but have not been mentioned in the secondary studies.
Hence, those quality attributes have zero context in which they were discussed.
Additionally, we list their six main quality attributes in Figure~\ref{fig:test_artifact_quality_model} in brackets.

Table~\ref{tab:qa_description} contains the descriptions of the 19 quality attributes identified in the secondary studies sorted according to Figure~\ref{fig:test_artifact_quality_model}. 
These descriptions of the quality attributes are direct quotes from the selected studies' authors.
Among those 19 quality attributes, \textit{Fault detection capability} and \textit{Completeness} are the most diverse ones with at least three different descriptions each, followed by \textit{Reusability}, \textit{Stability}, \textit{Fault detection effectiveness}, and \textit{Reliability} with two descriptions each.
It is worth to mention that the authors of the secondary studies did not mention any connection or influence of the ISO standard when describing the quality attributes.

\subsubsection{RQ2.3 -- Quality Measurements}\label{sec:rq.2.3.quality_measurements}

Together with the 30 quality attributes for test artifacts, we also found 100 unique quality measurements reported in the secondary studies for test artifact quality assessment.
In general, the most common quality measurements are \textit{coverage criteria} (mentioned in 67 unique contexts) while the other measurements appeared in at most seven contexts each.

In our study, we only reported the quality measurement - quality attributes associations which were explicitly mentioned in the secondary studies by the authors.
For example, in study S40 [21], the authors discussed several quality measurements (APFD, ASFD, TPFD, APFDc, NAPFD, RP, CE). They were described by the authors as metrics to evaluate test case prioritisation approaches.
There was no explicit test-artifact quality attribute connected to those quality measurements mentioned by the authors of S40.
Hence, we treated those quality measurements as stand-alone measurements.
Consequently, there are 82 out of the 100 identified quality measurements which had no explicitly stated connection to any particular quality attribute.
While they were reported as measurements of test artifact quality in general, the remaining 18 quality measurements were associated in the secondary studies with 11 specific quality attributes.
Those measurements are marked green in Figure~\ref{fig:test_artifact_quality_model}.
Also, as illustrated in Figure~\ref{fig:test_artifact_quality_model}, the largest set of quality measurements (seven out of 18 measurements) were for measuring the \textit{Fault detection capability} attribute.

Among those 18 quality measurements, \textit{coverage criteria} is the most frequent one.
Despite appearing in most contexts in form of various types, such as \textit{code coverage criteria}, \textit{model-flow coverage criteria}, \textit{script-flow coverage criteria}, \textit{data coverage criteria}, etc., \textit{coverage criteria} has been linked to quality attributes only in few contexts.
In particular, there are only seven contexts found in six secondary studies (S2, S8, S21, S23, S25, S32) where the authors mentioned that the \textit{coverage criteria} could be used to measure four identified quality attributes (\textit{Adequacy}, \textit{Completeness}, \textit{Fault detection capability}, and \textit{Fault detection effectiveness}).

There are only six out of those 18 quality measurements which had descriptions, as presented in Table~\ref{tab:QA_QMDescription}.
The six measurements are sorted according to Figure~\ref{fig:test_artifact_quality_model} and the descriptions originate from the secondary studies. 
Among those six measurements, \textit{coverage criteria} is the only one having multiple descriptions depending on the contexts and the quality attribute they are used to measure.

\subsection{RQ3 -- Quality attributes in the most common contexts}\label{sec:rq.3.quality_attribute_common context}
\begin{figure*}
\begin{center}
\includegraphics[width=1\textwidth]{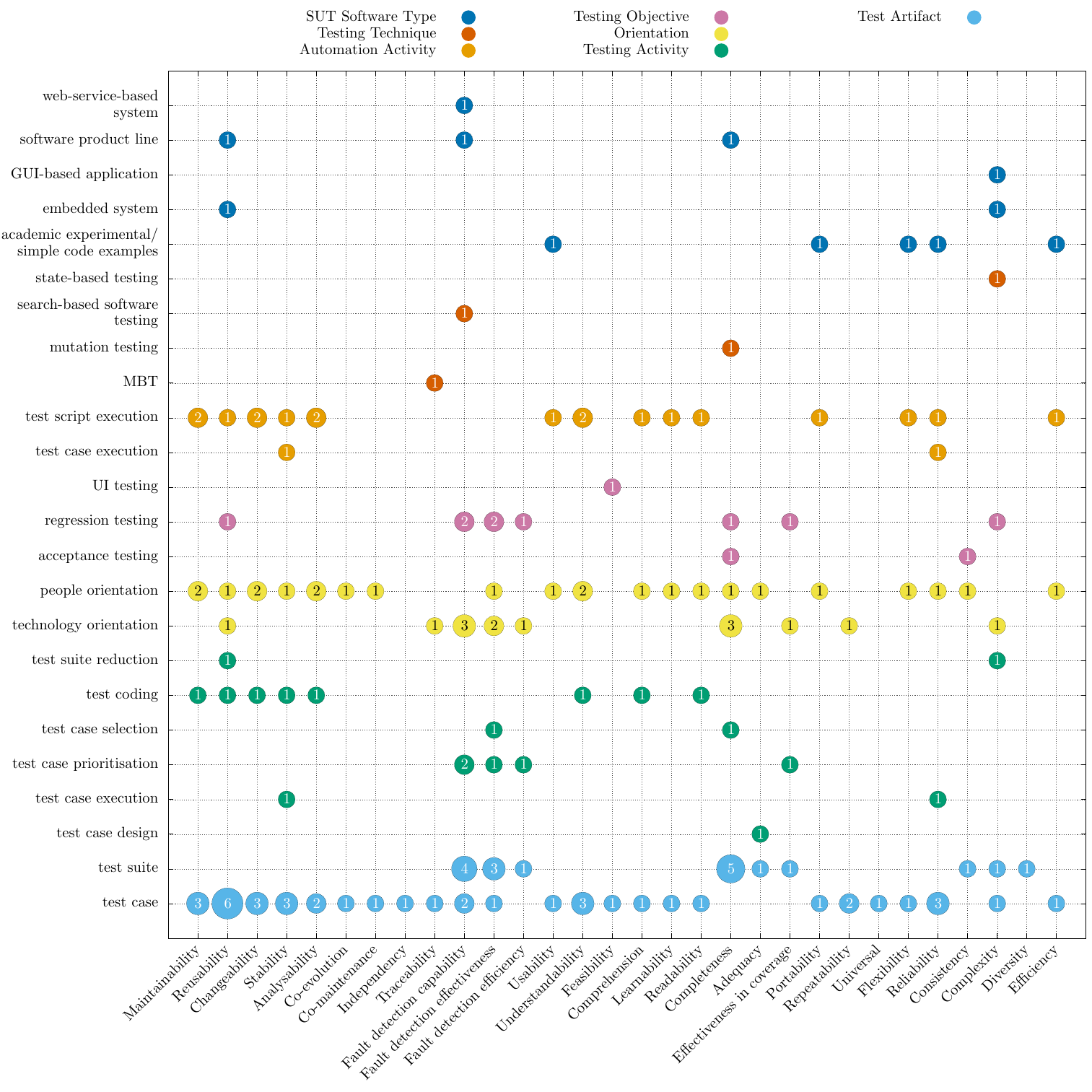}
\caption{Quality attributes in the most frequent contexts (The x-axis represents the quality attributes and the y-axis the contexts)}
\label{fig:context_QA}
\end{center}
\end{figure*}

As mentioned in Section~\ref{sec:rq2.1.context}, we identified 11 context dimensions describing testing-specific contexts where test artifact quality was discussed in the secondary studies.
Among the 11 context dimensions, there are seven dimensions which were discussed more frequently than others, namely \textit{Test artifact} (48 studies), \textit{Testing activity} (34 studies), \textit{Orientation} (33 studies), \textit{Testing objective} (20 studies), \textit{Automation activity} (17 studies), \textit{Testing technique} (15 studies) and \textit{SUT\_Software Type} (15 studies) (see Table~\ref{tab:context_frequency}).

In this section, we report the occurrences of the 30 identified quality attributes under the context values of those seven most frequent context dimensions.
An overview of the occurrences is illustrated in Figure~\ref{fig:context_QA}.
The horizontal axis represents the quality attributes sorted according to the number of their unique contexts, as shown in Figure~\ref{fig:test_artifact_quality_model}.
The vertical axis represents the context values of the most frequent context dimensions.
The context values without connected quality attributes were excluded from Figure~\ref{fig:context_QA}. Among those excluded context values, \textit{test case generation} is the most common context value under two dimensions, \textit{Testing Activity} and \textit{Automation activity}; \textit{safety-critical application} is the most common one under the \textit{SUT\_Software type} dimension.
Still, there is no quality attributes linked to those two context values.

Each number in Figure~\ref{fig:context_QA} shows how often a quality attribute has been studied.
For example, at the bottom-left corner of the figure, we can see that \textit{Maintainability} was studied under three contexts which contain the context value \textit{test case}.

Based on Table~\ref{tab:context_frequency}, we first compare context dimensions, which have been studied equally often to identify differences or similarities with regard to the thoroughness with which the connected quality attributes have been studied.
As a first case, we look at the two dimensions \textit{Testing activities} and \textit{Orientation}, which were discussed equally often.
From Figure~\ref{fig:context_QA}, we can see that there are 16 quality attributes connected to \textit{Testing activities} and 26 quality attributes connected to \textit{Orientation}.
By combining this with the data from Tables~\ref{tab:qa_description} and~\ref{tab:QA_QMDescription}, we find that, for the \textit{Testing activities} dimension, 75\% of the connected quality attributes had descriptions and 56\% of the connected quality attributes had information regarding their measurements.
In contrast, for the \textit{Orientation} dimension, only 58\% of the connected quality attributes had descriptions, and only 42\% of the connected quality attributes had information regarding their measurements.
Hence, we could see that the quality attributes under the \textit{Testing activities} dimension were studied more thoroughly than under the \textit{Orientation} dimension.

The above observation also applies when comparing the two context dimensions, \textit{Testing technique} and \textit{SUT\_Software type} in a similar manner.
Even though the dimensions were discussed equally often, the four quality attributes connected to \textit{Testing technique} were more well studied than the nine quality attributes connected to \textit{SUT\_Software type}.

\textit{Testing objective} and \textit{Automation activity} are two extreme cases. 
There are 14 quality attributes under the \textit{Automation activity} dimension, but only 36\% of them had descriptions, and 14\% of them had measurements information.
It is the dimension with the least informative quality attributes when comparing to the other dimensions.
Meanwhile, under the \textit{Testing objective} dimension, all nine connected quality attributes had descriptions, and 89\% of them had measurements detail.
It makes this dimension the one with the most well-studied quality attributes.

In a next step, we relate the quality attribute categories, as shown in Figure~\ref{fig:test_artifact_quality_model} to their coverage in different context dimensions, as can be derived from Figure~\ref{fig:context_QA}.
In this way, it is possible to identify context dimensions which disregard entire quality attribute categories.
Particularly, no quality attribute under the four categories, \textit{Maintainability}, \textit{Usability}, \textit{Portability}, and \textit{Reliability}, were connected to the \textit{Testing technique} dimension.
Likewise, no quality attributes under the two categories, \textit{Fault detection} and \textit{Coverage}, were connected to the \textit{Automation activity} dimension.
Also, the quality attributes under the category \textit{Portability} were not connected to two other dimensions, \textit{Testing activity} and \textit{Testing technique}.

Besides, for each quality attribute category, we identified the most commonly discussed quality attribute based on the number of context dimensions and context values the attribute was connected to, as shown in Figure~\ref{fig:context_QA}.
For example, \textit{Reusability} has been discussed in six out of seven context dimensions with 14 context values in total.
Those quality attributes are \textit{Reusability}, \textit{Fault detection capability}, \textit{Understandability}, \textit{Completeness}, \textit{Repeatability}, and \textit{Reliability}.
Nevertheless, there is no significant observation we could draw regarding the least popular quality attributes, namely \textit{Complexity}, \textit{Diversity}, and \textit{Efficiency} as they are the stand-alone attributes without any sub-attributes and were not discussed in many secondary studies (as described in Section~\ref{sec:rq.2.2.quality_attribute}).

\subsection{RQ4 -- Consensus on descriptions of quality attributes and quality measurements}\label{sec:rq.4.consensus}
In this section, we report our observations regarding the consensus on the descriptions of the identified quality attributes and quality measurements of test artifacts.
As mentioned in Section~\ref{sec:rq.2.2.quality_attribute}, there are 19 quality attributes which had descriptions in the secondary studies.
There are six quality attributes, namely \textit{Completeness}, \textit{Fault detection capability}, \textit{Fault detection effectiveness}, \textit{Reliability}, \textit{Reusability}, and \textit{Stability}, each of which had different descriptions (see Table~\ref{tab:qa_description}).

Our first observation is that we did not find any quality attribute with contradicting descriptions within the same context or in the same secondary study.
Specifically, with the above six quality attributes, even though each of them had different descriptions coming from different contexts, the descriptions are aligned.
Indeed, we found that they either represent different aspects of the quality attribute or are similar to some extent.
For example, the descriptions (2), (3), and (4) of the quality attribute \textit{Fault detection capability} in Table~\ref{tab:qa_description} all describe this attribute to be concerned with identifying defects or faults.
Furthermore, both descriptions (1) and (4) identify code coverage as important indicator of this quality attribute.
Likewise, with the quality attribute \textit{Reusability}, while one of its description was about reusing test cases between different testing levels, the other description was about another way of reusing test cases, which is reusing only some part of a test case.
Similarly, the quality attribute \textit{Reliability} has two descriptions covering two different but not contradictory aspects of a test failure, namely the frequency and the randomness. 

Our second observation is that the two quality attributes, \textit{Repeatability} and \textit{Universal}, share similarity in their descriptions.
Both attributes were described based on testing environments.
While the former attribute relies on the number of testing environment, the latter one is also based on test scenarios.

Our third observation is that there are five quality attributes with similarity in their measurements.
Those quality attributes are \textit{Adequacy}, \textit{Completeness}, \textit{Effectiveness in coverage}, \textit{Fault detection capability}, and \textit{Fault detection effectiveness}.
Even though their descriptions are not the same, the coverage criteria were the common way to quantify those attributes (as mentioned in their descriptions and measurements details).

\section{Discussion}\label{sec:discussion}
We discuss in this section the three main aspects from our results, covering the identified test artifact quality attributes, measurements and contexts.

\subsection{Test artifact quality attributes}\label{sec:discussion_qa}
We use ISO/IEC 25010:2011 to augment the test artifact quality attributes identified from the reviewed secondary studies with the standard's quality characteristic information, since (1) it is the most relevant and official source in the literature for quality models in software engineering, and (2) this idea has been implemented before~\cite{neukirchen_approach_2008,athanasiou_test_2014} but with the now withdrawn standard ISO/IEC 9126.

As reported in Section~\ref{sec:rq.2.2.quality_attribute}, we identified 30 quality attributes from the selected secondary studies.
When comparing our findings with the software quality characteristics in ISO/IEC 25010:2011, we found that there are 15 quality sub-characteristics, shown at the bottom of Figure~\ref{fig:test_artifact_quality_model} and belonging to six main quality characteristics shown in brackets, which could also be used to characterize test artifact quality. In Table~\ref{tab:qa_description_ISO} we propose descriptions, based on the definitions of the same quality characteristics in ISO/IEC 25010:2011, for the five out of six main quality characteristics (the main quality attribute \textit{reliability} is excluded as we found its descriptions in the literature) and 15 quality sub-characteristics for test artifacts that we did not identify in the secondary studies.

The reason for not finding those sub-characteristics in the selected secondary studies could be that they were not relevant to the studies' scopes, especially when most of the studies did not consider test artifact quality explicitly.
Another reason could be that the studies' syntheses and primary study details were not sufficiently detailed, preventing us to extract information about quality attributes (such as S31, according to our quality assessment's results, see Figure~\ref{fig:quality_score_sorted}).

Overall, as a combination of the 30 quality attributes from the selected secondary studies and the 20 additional quality attributes from the ISO standard, our quality model (see Figure~\ref{fig:test_artifact_quality_model}) contributes to researchers in two aspects.
First, it shows which quality attributes should be considered from the ISO standard when researchers characterize test artifact quality.
Second, it also shows what additional quality attributes (Table~\ref{tab:qa_description_ISO}) could be relevant when researchers characterize the test artifact quality based on the ISO standard. However, there are several aspects of test artifact quality that are not considered in the current model or cannot be solved by a static model alone. Test smells~\cite{van2001refactoring} are, similar to code smells~\cite{fowler2018refactoring}, indications that point to deeper, not immediately visible, issues in test code~\cite{garousi2018smells}. Spadini et al.~\cite{spadini2018relation} studied over 200 releases of 10 software systems and found that smelly test cases are more change- and defect prone (i.e. potentially less maintainable and reliable) and lead to more defect prone production code. Hence, a model of test smells with associated quality attributes could complement our test artifact quality model. 
We did not consider test data in the test artifact context dimension (see Table~\ref{tab:contextDimensions}) as we did not encounter any secondary studies that cover the topic of test \emph{data} quality in our review. While the practice of testing deep learning applications is still at an early stage~\cite{ma2018deepgauge}, the quality of test data for such systems is essential. Hence, extending the test quality model to include also means to define test data quality would be beneficial. Finally, the test quality model presented in this paper does not provide decision support for prioritizing or selecting attributes for particular test phases. Depending on development progress, product maturity, or even developer/tester experience, different aspects of test artifacts, and hence quality attributes, might need to be prioritized.

\begin{table*}
{    
    \footnotesize
    \begin{center}
    \caption{Descriptions of Test Artifact Quality Attributes (based on ISO/IEC 25010:2011)}
    \label{tab:qa_description_ISO}
   \begin{tabular}{p{0.18\textwidth}p{0.72\textwidth}}
	\toprule
	Quality Attribute	&	Description\\
	\midrule
	Satisfaction	&	Degree to which user needs are satisfied when a test artifact is used in a specified context of use  \\
	Usefulness	&	Degree to which a user is satisfied with their perceived achievement of pragmatic goals, including the results of use and the consequences of use   \\
	Trust	&	Degree to which a user or other stakeholder has confidence that a test artifact will behave as intended. \\
	Pleasure	&	Degree to which a user obtains pleasure from fulfilling their personal needs. Personal needs can include needs to acquire new knowledge and skills.   \\
	Functional suitability	&	Degree to which a test suite covers its given testing objectives which could be about fault detection, SUT performance evaluation, etc.  \\
	Functional completeness	&	Degree to which a test suite covers all the specified testing objectives. \\
	Functional correctness	&	Degree to which a test artifact provides the correct results with the needed degree of precision. \\
	Functional appropriateness	&	Degree to which a test case facilitates the accomplishment of specified tasks and objectives. For example, a user is only presented with the necessary steps to execute the test case, excluding any unnecessary steps.    \\
	Performance efficiency	&	Performance relative to the amount of resources used under stated conditions. Resources can include the software and hardware configuration of the system under test.   \\
	Time behaviour	&	Degree to which the response and processing times of a test artifact when executing.  \\
	Resource utilization	&	Degree to which the amounts and types of resources used by a test artifact when executing.    \\
	Compatibility	&	Degree to which a test artifact can exchange information with other test artifacts, and/or accomplish its required testing objective, while sharing the same hardware or software environment.    \\
	Co-existence	&	Degree to which a test artifact could accomplish its testing objective efficiently while sharing a common environment and resources with other test artifacts, without detrimental impact on any other test artifacts. For example, it could be between different test suites or test cases for running in parallel while using the same resources. \\
	Usability	&	Degree to which a test artifact can be used by specified users to achieve specified testing objectives with effectiveness, efficiency and satisfaction in a specified context of use. \\
	Appropriateness recognizability	&	Degree to which users can recognize whether a test artifact is appropriate for their needs e.g.: for test case selection in regression testing.    \\
	Operability	&	Degree to which a test script has attributes that make it easy to operate and control.    \\
	Maturity	&	Reflect on the number of reliable test cases in a test suite.   \\
	Availability	&	Degree to which a test artifact is operational and accessible when required for use.   \\
	Fault tolerance	&	Degree to which a test suite could accomplish its testing objective when its test case(s) fail due to hardware or software faults.   \\
	Recoverability	&	Degree to which, in the event of an interruption or a failure, a test suite can recover the data directly affected and re-establish its desired state. For example, if the tear-down of the failed test case is set up properly then the test suite could  re-establish its desired state then execute subsequent test cases.  \\
	\bottomrule
\end{tabular}

    \end{center}
}
\end{table*}

\subsection{Test artifact quality measurements}\label{sec:discussion_qm}
As reported in Section~\ref{sec:rq.2.3.quality_measurements}, we identified 100 unique quality measurements from the selected secondary studies.
Nonetheless, only 18 of those measurements were connected to quality attributes (11 in total).
Also, only six of those 18 measurements were thoroughly described in the secondary studies.
This means, in turn, that there were 19 quality attributes without measurement information and 12 quality measurements without description.
A possible explanation for not finding measurements for quality attributes and measurement definitions is that the information is not relevant to the secondary studies' scopes, hence not reported.
This lack of a consolidated view for measurements is a common issue in many different areas of software engineering as illustrated, for example, by Kitchenham~\cite{kitchenham2010s} with regard to software metrics and by Unterkalmsteiner et al.~\cite{unterkalmsteiner2011evaluation} with regard to measurements for software process improvement.
Similarly, there is still a need for a consolidated view of how to measure the quality attributes of test artifacts.

\subsection{Most frequently reported test artifact quality contexts}\label{sec:discussion_mostFrequentContext}
In Section~\ref{sec:rq.3.quality_attribute_common context}, we reported how often the 30 identified quality attributes have been studied in the seven most frequent contexts (illustrated in Figure~\ref{fig:context_QA}).
We found that some quality attributes were studied better, in terms of having descriptions and measurement information, in a certain context dimension than in another dimension despite that the dimensions were reported equally often.
Particularly, the quality attributes connected to the \textit{Testing activity}, \textit{Testing objective}, and \textit{Testing technique} dimensions were more studied than those connected to the \textit{Orientation}, \textit{Automation activity}, and \textit{SUT\_Software type} respectively.

Our second finding was that some context dimensions could not be associated with certain groups of quality attributes.
For example, no quality attribute under the four categories of \textit{Maintainability}, \textit{Usability}, \textit{Portability}, and \textit{Reliability} was associated with the \textit{Testing technique} dimension.
One possible explanation is that these context dimensions were not relevant to the scope of the related secondary studies and were therefore not discussed.
Nonetheless, one could argue that for high quality test artifacts, the majority of quality attributes are relevant, independent of context. The difficult question is however to find the optimal selection of attributes for a positive return on investment. Knowing all potential relevant quality attributes is a first step towards being able to make such a prioritization.

In the following, we discuss the potential implications of missing quality attributes for two frequently discussed contexts, viz., \textit{test case generation}, and \textit{regression testing}.

The context dimension \textit{test case generation} is the most extreme case in terms of missing quality attributes, since none have been discussed in this context. A possible explanation for this surprising finding is that these secondary studies focused on the comparison of different techniques for test case generation, reporting only quality measurements without associating them with quality attributes.
Nevertheless, this might indicate a need for further study, since quality attributes in the \textit{Fault Detection} and the \textit{Coverage} category likely need to be considered when discussing test artifacts.
Furthermore, it is conceivable that quality attributes, such as \textit{Analysability} and \textit{Traceability}, are also relevant in the context of \textit{test case generation}.
For example, while generated test cases might not need to be understandable if they are executed automatically, it still needs to be possible to trace them to the initial requirements.

In contrast to \textit{test case generation}, the \textit{regression testing} context is associated with the quality attribute categories \textit{Fault Detection} and \textit{Coverage}, which have been identified as relevant for test artifacts in general.
However, an important quality attribute, \textit{Reliability}, has not been discussed in the context of \textit{regression testing}. Thus, further investigation in this direction is needed.

Quality attributes have been discussed in many different contexts.
However, our systematic mapping of contexts and quality attributes has revealed gaps even in mature contexts, such as \textit{regression testing}.
Therefore, our mapping can serve as a basis for a more systematic exploration of quality attributes of test artifacts.

\subsection{Least frequently reported test artifact quality contexts}\label{sec:discussion_lestFrequentContext}
As reported in Section~\ref{sec:rq2.1.context}, we identified context dimensions which are potentially relevant for describing artifact quality in testing-specific contexts, but were not identified in the selected secondary studies.
The context dimensions are \textit{Automation level} in terms of \textit{Automation degree}, \textit{Quality type}, \textit{Quality standards}, and \textit{Solution/Improvement} in terms of \textit{correction}, \textit{detection} and \textit{prevention} (as shown in Table~\ref{tab:contextDimensions}).
While the Automation level dimension is based on our knowledge in software testing, the other dimensions were inspired by ISO/IEC 25010:2011 and the general concept of having a solution or improvement for test smells~\cite{garousi_systematic_2016}.
We argue that those dimensions are as important as the other dimensions for describing the context of test artifact quality for two reasons.
First, testing automation is a rising factor in software testing~\cite{garousi_when_2016} and the automation degree can provide a basis for prioritizing quality attributes.
Second, it is insufficient to describe only the characteristics of good-quality test artifacts without considering ways to detect, prevent, and correct bad quality in test artifacts.
This is important to ensure the creation of good-quality test artifacts.

Nonetheless, those context dimensions were not found in any selected secondary studies.
One explanation is that most of the selected secondary studies (except S31) did not focus on test artifact quality.
Hence, reporting context dimensions of test artifact quality might not have been deemed as necessary.
On top of that, none of the selected secondary studies mentioned the ISO/IEC standard for software quality as a reference for the information regarding test artifact quality in their reviews.
Regarding the secondary study S31, we found that the only mentioned context information was \textit{Test artifact} type (\textit{test cases} or \textit{test suite}). 
Also according to our quality assessment's results (see Figure~\ref{fig:quality_score_sorted}), the study did not fulfill the synthesis criterion (C3) and did not have sufficient details regarding their selected primary studies (C5) (see Table~\ref{tab:quality_assessment_criteria}).
Hence, it is possible that the three relevant dimensions were missed in S31.

\subsection{Research roadmap}
While this tertiary review uses a large base of existing literature to propose a test artifact quality model, gaps in this same literature need to be filled in order to complete the model.
Furthermore, specific instructions on how to instantiate the model in practice are missing. In this section, we briefly draw a path on how to fill these research gaps that would lead to a usable and evaluated test artifact model in practice.


\subsubsection{To complete the test-artifact quality model}
\paragraph{Context dimensions.}
Context dimensions are important to describe test-artifact quality in the software-testing specific context. 
However, there are still context dimensions which we could not collect information about in the literature (as mentioned in Section~\ref{sec:discussion_lestFrequentContext}).
Hence, we find it necessary to investigate the extent to which other context dimensions (such as Automation level in terms of Automation degree, Quality type, Quality standards, and Solution/Improvement in terms of correction, detection and prevention) could be used to describe test-artifact quality.

\paragraph{ISO-based quality attributes.}
We argued that there are several quality attributes from the ISO/IEC 25010:2011 (as mentioned in Section~\ref{sec:discussion_qa}) which could be used to characterize the test-artifact quality. 
However, we could not find relevant information regarding those quality attributes in the literature.
Hence, there is also a need to investigate the extent to which those extra quality attributes inspired by the ISO standard (as shown in Figure~\ref{fig:test_artifact_quality_model} and Table~\ref{tab:qa_description_ISO}) could be used to measure test-artifact quality and how to quantify those extra quality attributes.

\paragraph{Test-artifact quality in mature contexts.}
We also found that there are mature and common contexts in which quality attributes and measurements have not been discussed thoroughly as such "regression testing" and "test case generation" (as mentioned in Section~\ref{sec:discussion_mostFrequentContext}).
Hence, we also find it necessary to have a more systematic view of quality attributes and how to measure them in those contexts.

\paragraph{Test smells and test data.}
As test smells are closely connected to test-artifact quality and  test-data quality is also essential (as mentioned in Section~\ref{sec:discussion_qa}), we believe that there is a need to investigate on how to integrate test smells and test data into the current quality model.

\subsubsection{To instantiate the test-artifact quality model}
Once the quality model is complete, it is necessary to provide instruction on how to instantiate the model in specific contexts.
One important aspect would be quality attributes prioritisation.
Our hypothesis is that depending on different factors such as testing phases, development progress, product maturity, developer/tester experience, some quality attributes of test artifacts need to be prioritized and/or selected.
However, our current quality model does not capture this aspect of customisation.

\subsubsection{To evaluate the test-artifact quality model}
There is also a need to assess the quality of the complete model itself such as its performance, effectiveness at providing reliable measurements to detect bad-quality test cases, or its usefulness at supporting practitioners without researchers' active involvement.

\section{Conclusions and Future Work}\label{sec:conclusion}
Software testing continues to play an essential role in software development.  We rely on testing to gain confidence in the quality of new features and lack of presence of any regressions in existing functionality.  

The central artifacts in testing are the individual test cases and their collections. Therefore, defining and measuring the quality of these testing-artifacts is important for both research and practice. 

However, no comprehensive model for the quality of these artifacts was available. We address this gap in this tertiary study. We capitalized on the large number of available secondary studies on related topics and utilized ISO/IEC 25010:2011 to develop a quality model.

The quality model presented in this study can support:
\begin{itemize}
\item describing new guidelines and templates for designing new test cases.
\item developing assessment tools for evaluating existing test cases and suites.
\end{itemize}

To validate the model, we will systematically collect feedback from practitioners and academics. We also intend to use it to assess test artifacts together with our industrial collaborators. 

\section*{Acknowledgment}
This work has been supported by ELLIIT, a Strategic Area within IT and Mobile Communications, funded by the Swedish Government. The work has also been supported by research grant for the VITS project (reference number 20180127) from the Knowledge Foundation in Sweden.

\bibliographystyle{cas-model2-names}
\bibliography{mybib}

\newpage

\onecolumn

\begin{appendices}
\section{Quality Assessment Result}
\label{appendix:qualityAssessmentResult}
\begin{center}
{    
    \footnotesize
    \captionof{table}{Quality Assessment}
    \label{tab:quality_assessment_result}
   \begin{tabular}{cccccccc}
	\toprule
	ID	&	QA\_C1	&	QA\_C2	&	QA\_C3a	&	QA\_C3b	&	QA\_C4	&	QA\_C5	&	Quality Score	\\
	\midrule
	S1	&	1	&	0.5	&	1	&	0	&	0	&	1	&	3.5	\\
	S2	&	1	&	1	&	1	&	0	&	0	&	1	&	4	\\
	S3	&	1	&	1	&	1	&	0	&	0	&	1	&	4	\\
	S4	&	1	&	1	&	1	&	0	&	1	&	1	&	5	\\
	S5	&	1	&	0.5	&	0	&	0	&	1	&	0.5	&	3	\\
	S6	&	0.5	&	1	&	1	&	0	&	1	&	0.5	&	4	\\
	S7	&	0.5	&	1	&	1	&	0	&	0	&	1	&	3.5	\\
	S8	&	1	&	1	&	1	&	0	&	1	&	0.5	&	4.5	\\
	S9	&	1	&	1	&	1	&	0	&	0	&	0.5	&	3.5	\\
	S10	&	0	&	0.5	&	0	&	0	&	0	&	0.5	&	1	\\
	S11	&	1	&	0.5	&	0	&	0	&	0	&	1	&	2.5	\\
	S12	&	1	&	1	&	1	&	0	&	1	&	1	&	5	\\
	S13	&	1	&	1	&	1	&	0	&	0	&	1	&	4	\\
	S14	&	1	&	1	&	1	&	0	&	0.5	&	1	&	4.5	\\
	S15	&	1	&	1	&	1	&	0	&	0	&	0.5	&	3.5	\\
	S16	&	1	&	0.5	&	1	&	1	&	0.5	&	1	&	5	\\
	S17	&	1	&	0.5	&	1	&	0	&	0	&	0.5	&	3	\\
	S18	&	1	&	1	&	1	&	0	&	0	&	0.5	&	3.5	\\
	S19	&	1	&	1	&	1	&	0	&	0	&	1	&	4	\\
	S20	&	1	&	0.5	&	1	&	0	&	1	&	0.5	&	4	\\
	S21	&	1	&	0.5	&	1	&	0	&	0	&	1	&	3.5	\\
	S22	&	1	&	0.5	&	1	&	0	&	0	&	0.5	&	3	\\
	S23	&	1	&	0.5	&	0	&	0	&	1	&	0.5	&	3	\\
	S24	&	1	&	0.5	&	1	&	0	&	0	&	0.5	&	3	\\
	S25	&	1	&	1	&	1	&	0	&	0	&	1	&	4	\\
	S26	&	1	&	0.5	&	1	&	0	&	1	&	1	&	4.5	\\
	S27	&	1	&	1	&	1	&	0	&	1	&	0.5	&	4.5	\\
	S28	&	1	&	0.5	&	1	&	0	&	1	&	1	&	4.5	\\
	S29	&	1	&	1	&	1	&	0	&	0	&	0.5	&	3.5	\\
	S30	&	1	&	0.5	&	1	&	1	&	0.5	&	1	&	5	\\
	S31	&	1	&	0.5	&	0	&	0	&	0	&	0.5	&	2	\\
	S32	&	1	&	0.5	&	1	&	0	&	1	&	1	&	4.5	\\
	S33	&	1	&	0.5	&	1	&	0	&	0.5	&	1	&	4	\\
	S34	&	1	&	1	&	1	&	0	&	0	&	1	&	4	\\
	S35	&	1	&	0.5	&	1	&	0	&	0	&	1	&	3.5	\\
	S36	&	1	&	1	&	1	&	0	&	0	&	0.5	&	3.5	\\
	S37	&	1	&	0.5	&	1	&	0	&	0	&	0.5	&	3	\\
	S38	&	1	&	0.5	&	1	&	1	&	1	&	1	&	5.5	\\
	S39	&	1	&	0.5	&	1	&	0	&	1	&	0.5	&	4	\\
	S40	&	1	&	0.5	&	1	&	0	&	0	&	0.5	&	3	\\
	S41	&	1	&	0.5	&	1	&	0	&	1	&	0.5	&	4	\\
	S42	&	1	&	0.5	&	1	&	0	&	1	&	0.5	&	4	\\
	S43	&	1	&	0.5	&	1	&	0	&	1	&	0.5	&	4	\\
	S44	&	0	&	0.5	&	0	&	0	&	0	&	0.5	&	1	\\
	S45	&	1	&	0.5	&	1	&	0	&	0	&	0.5	&	3	\\
	S46	&	1	&	0.5	&	0	&	0	&	0	&	0.5	&	2	\\
	S47	&	1	&	1	&	1	&	0	&	0	&	1	&	4	\\
	S48	&	1	&	0.5	&	1	&	0	&	0	&	1	&	3.5	\\
	S49	&	1	&	1	&	1	&	0	&	0	&	1	&	4	\\
	\bottomrule
\end{tabular}
}
\end{center}

\newpage

\section{Publication venues of the 49 selected secondary studies}
\label{appendix:publicationVenues}

\begin{center}
{    
    \footnotesize
    \captionof{table}{Publication venues}
    \label{tab:pubVenues_1}
    \begin{tabular}{p{0.05\textwidth}p{0.5\textwidth}p{0.3\textwidth}}
    	\toprule
    	ID	&	Journal/Conference	&	Publisher	\\
    	\midrule
        S1	&	IEEE Access	&	IEEE	\\
    	S2	&	Journal of Systems and Software	&	Elsevier Inc.	\\
    	S3	&	Information and Software Technology	&	Elsevier B.V.	\\
    	S4	&	IEEE Access	&	IEEE	\\
    	S5	&	International Journal of Applied Engineering Research	&	Research India Publications	\\
    	S6	&	Proceedings of the ACM Symposium on Applied Computing	&	Association for Computing Machinery	\\
    	S7	&	Information and Software Technology	&	Elsevier B.V.	\\
    	S8	&	Information and Software Technology	&	Elsevier B.V.	\\
    	S9	&	Journal of Systems and Software	&	Elsevier Inc.	\\
    	S10	&	Proceedings - 41st Euromicro Conference on Software Engineering and Advanced Applications, SEAA 2015	&	IEEE	\\
    	S11	&	ICEIS 2018 - Proceedings of the 20th International Conference on Enterprise Information Systems	&	SciTePress	\\
    	S12	&	Information and Software Technology	&	Elsevier B.V.	\\
    	S13	&	International Journal on Software Tools for Technology Transfer	&	Springer Verlag	\\
    	S14	&	IEEE Transactions on Software Engineering	&	IEEE	\\
    	S15	&	Information and Software Technology	&	Elsevier B.V.	\\
    	S16	&	IEEE Access	&	IEEE	\\
    	S17	&	Information and Software Technology	&	Elsevier B.V.	\\
    	S18	&	Journal of Computer Science	&	Science Publications	\\
    	S19	&	IEEE Access	&	IEEE	\\
    	S20	&	Computacion y Sistemas	&	Instituto Politecnico Nacional	\\
    	S21	&	ACM Computing Surveys	&	Association for Computing Machinery	\\
    	S22	&	ESEM'08: Proceedings of the 2008 ACM-IEEE International Symposium on Empirical Software Engineering and Measurement	&	IEEE	\\
    	S23	&	Journal of Telecommunication, Electronic and Computer Engineering	&	Universiti Teknikal Malaysia Melaka	\\
    	S24	&	Proceedings - 42nd Euromicro Conference on Software Engineering and Advanced Applications, SEAA 2016	&	IEEE	\\
    	S25	&	IEEE Transactions on Reliability	&	IEEE	\\
    	S26	&	Software Quality Journal	&	Springer New York LLC	\\
    	S27	&	Computer Science Review	&	Elsevier Ireland Ltd	\\
    	S28	&	Information and Software Technology	&	Elsevier B.V.	\\
    	S29	&	Empirical Software Engineering	&	Kluwer Academic Publishers	\\
    	S30	&	Information and Software Technology	&	Elsevier B.V.	\\
    	S31	&	Proceedings - 9th Knowledge Management International Conference (KMICe)	&	UNIV UTARA MALAYSIA PRESS	\\
    	S32	&	ACM Computing Surveys	&	Association for Computing Machinery	\\
    	S33	&	Journal of Systems and Software	&	Elsevier Inc.	\\
    	S34	&	Information and Software Technology	&	Elsevier B.V.	\\
    	S35	&	ACM Computing Surveys	&	Association for Computing Machinery	\\
    	S36	&	Proceedings - 2015 IEEE World Congress on Services, SERVICES 2015	&	IEEE	\\
    	S37	&	Informatica (Slovenia)	&	Slovenian Society INFORMATIKA	\\
    	S38	&	Information and Software Technology	&	Elsevier B.V.	\\
    	S39	&	Information and Software Technology	&	Elsevier B.V.	\\
    	S40	&	Software Quality Journal	&	Springer New York LLC	\\
    	S41	&	ACM International Conference Proceeding Series	&	Association for Computing Machinery	\\
    	S42	&	International Journal of Software Engineering and Knowledge Engineering	&	World Scientific Publishing Co. Pte Ltd	\\
    	S43	&	ACM International Conference Proceeding Series	&	Association for Computing Machinery	\\
    	S44	&	International Journal of Advanced Computer Research	&	Accent Social and Welfare Society	\\
    	S45	&	Applied Soft Computing Journal	&	Elsevier Ltd	\\
    	S46	&	Proceedings - Asia-Pacific Software Engineering Conference, APSEC	&	IEEE	\\
    	S47	&	Proceedings - 41st Euromicro Conference on Software Engineering and Advanced Applications, SEAA 2015	&	IEEE	\\
    	S48	&	Journal of Systems and Software	&	Elsevier Inc.	\\
    	S49	&	Information and Software Technology	&	Elsevier B.V.	\\
	    \bottomrule
    \end{tabular}
    }
\end{center}

\end{appendices}

\end{document}